\documentclass[journal]{IEEEtran}
\usepackage{cite}

\ifCLASSINFOpdf
   \usepackage[pdftex]{graphicx}
\else
\fi
\usepackage[cmex10]{amsmath}
\usepackage{amssymb}

\usepackage{algorithmic}
\usepackage{algorithm}

\usepackage{array}

\usepackage{stfloats}
\newtheorem{theorem}{Theorem}

\newcommand{\qed}{\nobreak \ifvmode \relax \else
      \ifdim\lastskip<1.5em \hskip-\lastskip
      \hskip1.5em plus0em minus0.5em \fi \nobreak
      \vrule height0.75em width0.5em depth0.25em\fi}

\begin{document}
%
\title{Spectrum Sensing with Small-Sized Datasets in Cognitive Radio: Algorithms and Analysis}
%
%
%

\author{Feng~Lin,~\IEEEmembership{Student Member,~IEEE,}
        Robert~C.~Qiu,~\IEEEmembership{Senior Member,~IEEE,}
				and James~P.~Browning,~\IEEEmembership{Member,~IEEE}

\thanks{Manuscript received xxxxx xx, 2013; revised xxxxx xx, 2014; accepted xxxxx xx, 2014. This work is funded by the National Science Foundation through three grants (ECCS-0901420, ECCS-0821658, and CNS-1247778), and the Office of Naval Research through two grants (N00010-10-1-0810 and N00014-11-1-0006).}
\thanks{Copyright (c) 2013 IEEE. Personal use of this material is permitted. However, permission to use this material for any other purposes must be obtained from the IEEE by sending a request to pubs-permissions@ieee.org.}
\thanks{F. Lin and R. C. Qiu are with the Department of Electrical and Computer Engineering, Center for Manufacturing Research, Tennessee Tech University, Cookeville, TN, 38505, USA,
(e-mail: fenglin@ieee.org; rqiu@tntech.edu).}
\thanks{J. P. Browning is with Sensors Directorate, AFRL/RYMD, Wright Patterson AFB, Dayton, OH, 45433, USA, (e-mail: James.Browning@wpafb.af.mil).}
}

\maketitle

\begin{abstract}
Spectrum sensing is a fundamental component of cognitive radio. How to promptly sense the presence of primary users is a key issue to a cognitive radio network. The time requirement is critical in that violating it will cause harmful interference to the primary user, leading to a system-wide failure. The motivation of our work is to provide an effective spectrum sensing method to detect primary users as soon as possible. In the language of streaming based real-time data processing, short-time means small-sized data. In this paper, we propose a cumulative spectrum sensing method dealing with limited sized data. A novel method of covariance matrix estimation is utilized to approximate the true covariance matrix. The theoretical analysis is derived based on McDiarmid's concentration inequalities and random matrix theory to support the claims of detection performance. Comparisons between the proposed method and other traditional approaches, judged by the simulation using a captured digital TV signal, show that this proposed method can operate either using smaller-sized data or working under lower SNR environment.
\end{abstract}


%
\IEEEpeerreviewmaketitle

\section{Introduction}
%
%
%
%
\IEEEPARstart{A}{s} a limited natural resource, wireless spectrum becomes increasingly scarce due to the evolution of various wireless technologies. However, it is not utilized efficiently; the current utilization of a licensed spectrum varies from 15\% to 85\%~\cite{force2002spectrum}. The number is even lower in rural areas. Cognitive radio (CR) is a key technology to mitigate the overcrowding of spectrum space based on its capability to perform dynamic spectrum access (DSA). When a primary user (PU) starts its transmission, the secondary user (SU) must vacate the frequency band as soon as possible. SUs that fail to sense the occupied spectrum and vacate the spectrum in time will cause unexpected harmful interference to PUs and even damage the whole cognitive radio network. With SUs frequently moving between regions with different densities of PUs, such as in vehicular applications, rapid PU detection is of great importance. Standards exist to address this detection requirement. For example, the IEEE 802.22 standard for unlicensed operation in the TV band regulates that PUs should be detected within 2 seconds of their appearance~\cite{kim2007experimental}.

As we have seen, it is clear that one fundamental requirement of a CR system is for SUs to use spectrum sensing to find spectral holes. Each SU should be able to sense a PU's existence accurately to avoid interference, even when the PU's signal is weak. Looking at it in this light can lead one to see how spectrum sensing could be treated as a signal detection problem. There has been plenty of research on spectrum sensing using classical detection schemes, such as energy detection~\cite{son1992, 997120, 1204119}, matched filter detection~\cite{4840525, 4840526, Zeng:2010}, cyclostationary feature detection~\cite{dandawate1994statistical, 4413137, 3769}, and covariance matrix based detection~\cite{Zeng2007, 5089517, 4698617, 4221495}. Covariance matrix related spectrum sensing algorithms were extended by employing multiple antennas at the cognitive receiver~\cite{Wang2010}. A suboptimal multi-antenna detector under unknown noise has also been proposed~\cite{lopez2010multiantenna}. Feature template matching (FTM)~\cite{Zhang} extracts signal features as the leading eigenvector of signal's covariance matrix. The feature is stable over time for non-white wide-sense stationary (WSS) signals while random for white noise~\cite{Zhanga}. Kernel feature template matching (KFTM)~\cite{Hou2011} extended the linear FTM to a nonlinear FTM by mapping data from a input space to a high dimensional feature space. The mapping is implemented by the so-called kernel trick. Applications of kernel-based learning in cognitive radio network have been proposed in literature~\cite{ding2013kernel}, the algorithms after kernel mapping have gained significant performance improvements over their linear counterparts at the price of generally higher computational complexity. Generalized function of matrix detection (FMD) has been employed for spectrum sensing, through the use of the function of random matrix and matrix inequality~\cite{Feng2012comml, Feng2012wdd, qiu2013cognitive}. A two-dimensional sensing framework has been proposed for spatial-temporal opportunity detection in cognitive radio~\cite{6409508}, which exploits correlations in time and space simultaneously by fusing sensing results in a spatial-temporal sensing window. A selective-relay-based cooperative sensing scheme has been proposed for both the spectrum sensing and secondary transmissions to achieve a reliable and efficient cognitive radio system~\cite{zou2012cooperative}, in which a dedicated channel usually used for reporting initial detection results for fusion is not essentially needed.

The vast majority of the research on spectrum sensing required large-sized datasets for processing to make a final decision. It was difficult to solve the sensing problem with limited received signal samples under low SNR. To circumvent this difficulty, we try to explore and utilize the core idea of quickest detection. Quickest detection~\cite{quickestbook} tries to detect the change of two different random processes with the shortest delay. If the change happens at the beginning of spectrum sensing, the goal of quickest detection is similar to that of sequential detection. The successive refinement algorithm was proposed which combined both the generalized likelihood ratio (GLR) and parallel cumulative sum (CUSUM) tests for quickest spectrum sensing~\cite{li2008quickest}. This algorithm used only average run lengths to measure the detection delay performance without the results on detection probability performance. Collaborative quickest spectrum sensing via random broadcast was investigated by first deriving a necessary condition for the optimal broadcast probability via asymptotic and variation analysis, then proposing a threshold broadcast scheme~\cite{li2010collaborative}. This algorithm used ROC curve of average detection delay and false alarm rate as the performance metric, but no ROC curve of detection probability and false alarm rate was provided. Besides, this reference did not consider the impact of the change of SNR. A sequential change detection framework for quickest detection has also been established for cognitive radio systems~\cite{lai2008quickest}. A hidden markov model (HMM) for quickest detection has been proposed for spectrum sensing, and the effectiveness has been verified by the experimental tests using industrial standard wireless communications signal~\cite{chen2009quickest}. However, this reference failed to show the superiority of the proposed algorithm over other algorithms. Linear-based CUSUM statistics for different cooperative sensing scenarios with unknown parameters of the distribution have also been investigated in~\cite{zarrin2009cooperative}. A fast spectrum sensing algorithm based on the discrete wavelet packet transform has been proposed in~\cite{youn2006fast}. However, this algorithm focused only on the coarse detection and needed a fine detection stage to complete the whole spectrum sensing.

In this paper, we propose a cumulative spectrum sensing method with small datasets. This method works as a real-time processing of streaming data; every incoming sample generates a metric value to compare with the threshold in real-time, thus minimizing the sensing latency. The input data are first collected by the system, after which this sequential data is transformed into a sample covariance matrix. Because the size of the datasets is too small for calculating an accurate sample covariance matrix, oracle-approximating shrinkage estimation is utilized to get an accurate estimate of the true covariance matrix. The cumulative average is then incorporated to smooth the detection metric. Two performance metrics, the number of sample data needed for detection versus SNR and detection probability versus SNR, are used to support the superiority of the cumulative spectrum sensing method.  

The contributions of this paper are as follows.

\begin{enumerate}
	\item A novel cumulative spectrum sensing method has been proposed for detection with small-sized datasets. This method is efficient and effective in practice in that it can blindly detect the PU's signal without knowing any information about the noise or the PU's signal. Meanwhile, this method can work in a relatively low SNR environment given that the sampling period is limited. In other words, the total sample data is fixed.
	 
	\item The analysis based on McDiarmid's concentration inequalities of statistics has been performed to demonstrate the superiority and effectiveness of the detection performance. The statistics on both hypotheses have been found to converge to distinct constant mean values. As a result, the signal and noise can be distinguished.
	
	\item A threshold based on the false alarm probability has been proven to be stable without noise uncertainty problem. The probability distribution function of the test statistic under hypothesis $\mathcal{H}_{0}$ has been derived, which is found approximately to be a Gaussian distribution.
	
\end{enumerate}

The organization of this paper is as follows: in Section~\ref{sys}, a system model based on a binary hypothesis test and sample covariance matrix calculation is described; in Section~\ref{quick}, the cumulative spectrum sensing method with small datasets is presented with the corresponding mathematical foundations; Section~\ref{analy} presents a performance analysis including concentration inequalities of statistics and robust threshold; Section~\ref{simu} provides the numerical results using a DTV signal, and comparisons are made with other popular detection methods.

\textit{Notation}: In the following, we depict vectors in lowercase boldface letters and matrices in uppercase boldface letters. $\left ( \cdot  \right )^{T}$ means the transpose operator and $\mathrm{Tr}\left ( \cdot  \right )$ means the trace operator. $\left \| \cdot  \right \|_{F}$ represents the Frobenius norm and $\mathrm{inf}\left ( \cdot  \right )$ represents the infimum. 

%
 %
\section{System Model}
\label{sys}
\subsection{Binary Hypothesis Test}
In a secondary network, we consider each SU with one receive antenna to detect one PU's signal based on its own observation. Let $x(t)$ be the continuous-time received signal after unknown channel. Let $T_{s}$ be the sampling period. The received signal sample is 
\begin{equation}
\label{sample}
x\left [ n \right ]=x\left ( nT_{s} \right )
\end{equation}
There are two hypotheses to detect PU's signal's existence, $\mathcal{H}_{0}$, only the noise (no PU's signal) exists; and $\mathcal{H}_{1}$, both the PU's signal and the noise exist. The received signal samples under the two hypotheses are given respectively as follows:
\begin{equation}
\label{H_0}
\mathcal{H}_{0}:\! x\left [ n \right ]=w\left [ n \right ]
\end{equation}
\begin{equation}
\label{H_1}
\mathcal{H}_{1}:\! x\left [ n \right ]=s\left [ n \right ] + w\left [ n \right ]
\end{equation}
where $w\left [ n \right ]$ is the received white Gaussian noise, and each sample of $w\left [ n \right ]$ is assumed to be independent identical distribution (i.i.d.), with zero mean and variance $\sigma _{n}^{2}$. $s\left [ n \right ]$ is the received PU's signal samples after unknown channel with unknown signal distribution.
Though in practice, the noise $w\left [ n \right ]$ after analog-to-digital (ADC) is usually non-white, we can use pre-whitening techniques to whiten the noise samples. In the rest of this paper, the noise is considered white.

Two probabilities of interest are used to evaluate detection performance. One is detection probability $P_{d}$, that is, at hypothesis $\mathcal{H}_{1}$, the probability having detected the PU's signal. The other is false alarm probability $P_{fa}$, the probability having detected the PU's signal at hypothesis $\mathcal{H}_{0}$. Apparently, we want to obtain a high $P_{d}$ and a low $P_{fa}$. The requirements of $P_{d}$ and $P_{fa}$ depend on applications.

\subsection{Sample Covariance Matrix}
\begin{figure}[!t]
\centering
\includegraphics[width=3.4in]{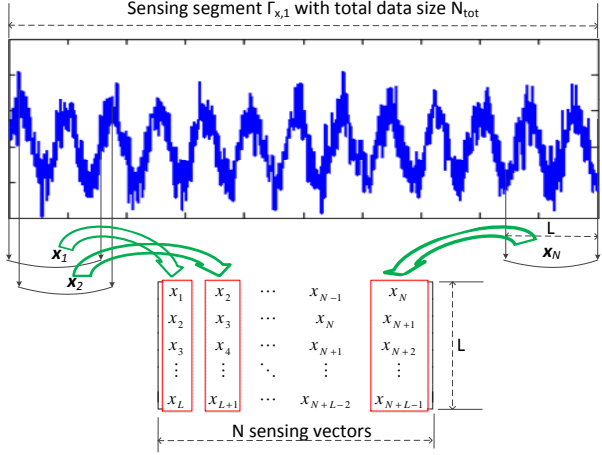}
\caption{Illustration of sensing segment and sensing vectors}
\label{segmentdraw}
\end{figure}

Assume spectrum sensing is performed based on the statistics of the $i^{th}$ sensing segment $\mathbf{\Gamma} _{x,i}$, which consists of $N_{tot}$ total sample data. The sensing segment $\mathbf{\Gamma} _{x,i}$ can be formed as $N$ sensing vectors with $L$ (called ``smoothing factor") consecutive output samples in each vector:
\begin{equation}
\label{segment}
\mathbf{\Gamma} _{x,i}=\left \{ \mathbf{x}_{\left ( i-1 \right )N+1},\mathbf{x}_{\left ( i-1 \right )N+2},\cdots ,\mathbf{x}_{\left ( i-1 \right )N+N} \right \}
\end{equation}
\begin{equation}
\label{vector_r}
\mathbf{x}_{i}=\left [ x\left [ i \right ],x\left [ i+1 \right ],\cdots ,x\left [ i+L-1 \right ] \right ]^{T}
\end{equation}
where $\mathbf{x}_{i}\sim \mathcal{N}\left ( \mathbf{\overline{x}}, \mathbf{R}_{x}\right )$, $\mathbf{\overline{x}}$ is the mean of $\mathbf{x}_{i}$ and $\mathbf{R}_{x}$ is covariance matrix of $\mathbf{x}_{i}$. Thus we have the equation 
\begin{equation}
\label{size}
N_{tot} = N + L - 1
\end{equation}
Here, to distinguish $N_{tot}$ and $N$, $N_{tot}$ is named total data size, and $N$ is named sample size (the number of columns of sensing segment matrix). $\mathbf{\Gamma} _{s,i}$ and $\mathbf{s}_{i}$ are defined in the same way as $\mathbf{\Gamma} _{x,i}$ and $\mathbf{x}_{i}$. The graphical illustration of sensing segment and sensing vector is provided in Fig.~\ref{segmentdraw} to facilitate the understanding of how to form the data matrix.

In practice, the covariance matrix of the observed signals is unknown. Thus, the unstructured classical estimator of $\mathbf{R}_{x}$, the sample covariance matrix, is adopted and defined as
\begin{equation}
\label{cov_mat}
\hat{\mathbf{R}}_{x}=\frac{1}{N}\sum_{i=1}^{N}\left (\mathbf{x}_{i}- \mathbf{\overline{x}} \right )\left (\mathbf{x}_{i} -\mathbf{\overline{x}} \right )^{T}
\end{equation}
Here, we assume the sample mean to be zero,
\begin{equation}
\label{cov_mat2}
\mathbf{\overline{x}}=\frac{1}{N}\sum_{i=1}^{N}\mathbf{x}_{i} = \mathbf{0}
\end{equation}
Then, the sample covariance matrix is simplified as
\begin{equation}
\label{cov_mat3}
\hat{\mathbf{R}}_{x}=\frac{1}{N}\sum_{i=1}^{N}\mathbf{x}_{i}\mathbf{x}_{i}^{T}
\end{equation}

Large sample work in multivariate analysis has traditionally assumed that $\frac{N}{L}$, the number of observations per variable, is large. Today, it is common for $L$ to be large or even huge, and so $\frac{N}{L}$ may be moderate to small and in extreme cases less than one~\cite{JCM_qiu2012}. In such case, an appropriate covariance matrix estimation is essentially needed.


\section{Spectrum Sensing with Small-Sized Datasets}
\label{quick}
Detecting the presence of PU's signal promptly is the basis of cognitive radio network. As soon as the PU is detected in the home channel, the SU has to vacate its home channel immediately. It is apparent the time requirement for cognitive radio system is of crucial importance to avoid interference. As a result, spectrum sensing should be performed in a short time. As the sampling rate is fixed for SU, sampling in a short time will only generate limited total size data $N_{tot}$. In some sense, detection in short time is equivalent to detection using small-sized datasets. In the following, we will focus more on the data sample size than the detection time, since the sample size is directly involved into the algorithm design. 

\subsection{Algorithm Fundamentals}
When $N_{tot}$ is small, the sample size $N$ would be comparable to matrix dimension $L$, even $N < L$. In such case, the sample covariance matrix $\hat{\mathbf{R}}_{x}$ is known to be a poor estimator of $\mathbf{R}_{x}$, which cannot describe the accurate statistical relationship within each sample. Many shrinkage estimators have been proposed under different performance measures by minimizing the mean-squared error (MSE) to approximate the true covariance matrix. In oracle-approximating shrinkage (OAS) estimation~\cite{robust_shrinkage}, the estimator $\hat{\mathbf{\Sigma }}$ is a trade-off between low bias and low variance, which is the solution to
\begin{equation}
\label{mini}
\begin{array}{l}
{\min _\rho }\quad {\rm{E}}\left\{ {\left\| {{\bf{\hat \Sigma }} - {{\bf{R}}_x}} \right\|_F^2} \right\}\\
{\rm{s}}{\rm{.}}{\kern 1pt} {\rm{t}}{\rm{.}}\quad {\bf{\hat \Sigma }} = \left( {1 - \rho } \right){{{\bf{\hat R}}}_x} + \rho {\bf{\hat F}}
\end{array}
\end{equation}
where $\hat{\mathbf{R}}_{x}$ is the sample covariance matrix defined in Eq.~(\ref{cov_mat}). The matrix $\hat{\mathbf{F}}$ is referred to as the shrinkage target, defined as
\begin{equation}
\label{F}
\hat{\mathbf{F}}=\frac{\mathrm{Tr}\left ( \hat{\mathbf{R}}_{x} \right )}{L}\mathbf{I}
\end{equation}
where $\mathbf{I}$ is a $L$ dimensional unitary matrix. Shrinkage coefficient $\rho$, usually between $0$ and $1$, is aimed at minimizing the MSE. The solution is shown in Theorem~\ref{theorem1} and the immediately followed iterations~\cite{chen2010shrinkage}.

\begin{theorem}
\label{theorem1}
Let $\hat{\mathbf{R}}_{x}$ be the sample covariance of a set of $L$-dimensional vectors $\left \{ \mathbf{x}_{i} \right \}_{i=1}^{n}$. If $\left \{ \mathbf{x}_{i} \right \}_{i=1}^{n}$ are i.i.d. Gaussian vectors with covariance $\mathbf{R}_{x}$, then the solution to \eqref{mini} is
\begin{equation}
\label{iter0}
\rho _{\mathrm{O}}=\frac{\left ( 1-\frac{2}{L} \right )\mathrm{Tr}\left (  \mathbf{R}_{x}^{2} \right )+\mathrm{Tr}^{2}\left ( \mathbf{R}_{x} \right )}{\left ( N+1-\frac{2}{L} \right )\mathrm{Tr}\left ( \mathbf{R}_{x}^{2} \right )+\left ( 1-\frac{N}{L} \right )\mathrm{Tr}^{2}\left ( \mathbf{R}_{x} \right )}
\end{equation}
\end{theorem}

Since $\mathbf{R}_{x}$ is hard to obtain, the OAS estimator is trying to approximate the solution in Eq.~(\ref{iter0}) via an iterative procedure. It initializes iterations with an initial guess of $\mathbf{R}_{x}$ and iteratively refine it. The iteration procedure is continued until convergence, which is
\begin{equation}
\label{iter1}
\hat{\rho} _{j+1}=\frac{\left ( 1-\frac{2}{L} \right )\mathrm{Tr}\left ( \hat{\mathbf{\Sigma}}_{j} \hat{\mathbf{R}}_{x} \right )+\mathrm{Tr}^{2}\left ( \hat{\mathbf{\Sigma}}_{j} \right )}{\left ( N+1-\frac{2}{L} \right )\mathrm{Tr}\left ( \hat{\mathbf{\Sigma}}_{j} \hat{\mathbf{R}}_{x} \right )+\left ( 1-\frac{N}{L} \right )\mathrm{Tr}^{2}\left ( \hat{\mathbf{\Sigma}}_{j} \right )}
\end{equation}
\begin{equation}
\label{iter2}
\hat{\mathbf{\Sigma }}_{j+1}=\left (1-\hat{\rho}_{j+1}   \right ) \hat{\mathbf{R}}_{x} + \hat{\rho}_{j+1} \hat{\mathbf{F}}
\end{equation}
The initial guess $\hat{\mathbf{\Sigma }}_{0}$ could be the sample covariance matrix $\hat{\mathbf{R}}_{x}$. The initial $\hat{\rho}_{0}$ could be any value between $0$ and $1$. Here, $\mathrm{Tr}( \hat{\mathbf{\Sigma}}_{j}^{2} )$ is replaced by $\mathrm{Tr}(\hat{\mathbf{\Sigma}}_{j} \hat{\mathbf{R}}_{x}  )$ since $\mathrm{Tr}( \hat{\mathbf{\Sigma}}_{j}^{2} )$ would always force $\hat{\rho} _{j}$ to converge to 1 while $\mathrm{Tr} ( \hat{\mathbf{\Sigma}}_{j} \hat{\mathbf{R}}_{x} )$ is not.

When the above iteration converges, we can get the following estimation:
\begin{equation}
\label{rho}
\hat{\rho} _{\mathrm{OAS}}= \mathrm{min}\left ( \frac{\left ( 1-\frac{2}{L} \right )\mathrm{Tr}\left ( \hat{\mathbf{R}}_{x}^{2} \right )+\mathrm{Tr}^{2}\left ( \hat{\mathbf{R}}_{x} \right )}{\left ( N+1-\frac{2}{L} \right )\left [ \mathrm{Tr}\left ( \hat{\mathbf{R}}_{x}^{2} \right )-\frac{\mathrm{Tr}^{2}\left ( \hat{\mathbf{R}}_{x} \right)}{L} \right ]},1 \right )
\end{equation}
In addition, $0< \hat{\rho} _{\mathrm{OAS}}< 1$. 

After using $\hat{\rho} _{\mathrm{OAS}}$ to substitute $\rho $ in \eqref{mini}, we can get the estimated covariance matrix as
\begin{equation}
\label{estcov}
\hat{\mathbf{\Sigma }}_{\mathrm{OAS}}=\left (1-\hat{\rho}_{\mathrm{OAS}}   \right ) \hat{\mathbf{R}}_{x} + \hat{\rho}_{\mathrm{OAS}} \hat{\mathbf{F}}
\end{equation}

Eigenvalues of sample covariance matrix are widely used in detection. The maximum-minimum eigenvalue (MME) algorithm~\cite{Zeng2007} uses the ratio of maximum and minimum eigenvalue, obtained from sample covariance matrix, as the detection metric. However, if the data size is not huge enough, the detection performance of MME will be compromised. It is natural to replace the eigenvalues from the sample covariance matrix with the eigenvalues from the estimated covariance matrix. The eigenvalues decomposed from $\hat{\mathbf{\Sigma }}_{\mathrm{OAS}}$ are denoted as: $\lambda _{1}\geq \cdots \geq \lambda _{L}$. The maximum eignevalue $\lambda _{1}$ and minimum eigenvalue $\lambda_{L}$ are defined as
\begin{equation}
\label{max}
{\lambda _1} = \mathop {\max }\limits_{{{\mathbf{u}}^H}{\mathbf{u}} = 1} {{\mathbf{u}}^H}{\widehat {\mathbf{\Sigma }}_{{\text{OAS}}}}{\mathbf{u}}
\end{equation}
\begin{equation}
\label{min}
{\lambda _L} = \mathop {\max }\limits_{\dim (\psi ) = L} \mathop {\min }\limits_{\mathop {{\mathbf{u}} \in \psi }\limits_{{{\mathbf{u}}^H}{\mathbf{u}} = 1} } {{\mathbf{u}}^H}{\widehat {\mathbf{\Sigma }}_{{\text{OAS}}}}{\mathbf{u}}
\end{equation}
where $dim(\psi)$ denotes the dimension of the subspace $\psi$.
 
The ratio of maximum and minimum eigenvalue can be calculated as
\begin{equation}
\label{mme}
T=\frac{\lambda _{1}}{\lambda _{L}}
\end{equation}

The CUSUM test~\cite{page1954continuous} is the optimal solution for minimizing delay and the central algorithm of non-Bayesian quickest detection, which requires the perfect knowledge of the distribution. In this paper, the distribution parameters of $\mathcal{H}_{1}$ are unknown. However, the fundamental idea of CUSUM test can be simplified and utilized here.

The stopping time for detecting the change is defined by
\begin{equation}
\label{stop}
t^{stop}=\mathrm{inf}\left ( t\mid Q_{t}\geq \gamma  \right )
\end{equation}
where the $\gamma$ is the detection threshold and $Q_{t}$ is the detection statistic at time slot $t$.
As we have discretized the time serials data into digital samples in Eq.~(\ref{sample}), similar to the above formula, the stopping total data samples for detecting the presence of PU's signal is given by
\begin{equation}
\label{stop1}
N_{tot}^{stop}=\mathrm{inf}\left ( N_{stop}\mid Q_{N_{stop}}\geq \gamma  \right ) + L -1
\end{equation}
Here, $Q_{N_{stop}}$ is the metric for detection when $N_{stop}$ sample size are involved and defined as 
\begin{equation}
\label{Metr}
Q_{N_{stop}}=\frac{q_{N_{stop}}}{N_{stop}}
\end{equation}
In some cases, the environment is so harsh that all the received data will be processed for the spectrum sensing decision, so $N_{stop}$ simply equals to $N$, which leads to
\begin{equation}
\label{Metr2}
Q_{N}=\frac{q_{N}}{N}
\end{equation}
$q_{\tau }$ can be computed recursively:
\begin{equation}
\label{stop2}
q_{\tau }=\mathrm{max}\left ( q_{\tau -1} + T_{\tau },0 \right ), 1\leq \tau\leq N
\end{equation}
where $T_{\tau }$ is defined by \eqref{mme} when $\tau$ sample size is used for covariance matrix calculation.
The initial value of $q_{0}$ is 0.

We noticed that $T_{\tau }$ is positive under both hypotheses, thus \eqref{stop2} is equivalent to
\begin{equation}
\label{stop3}
q_{\tau }=\sum_{k=1}^{\tau }T_{k}
\end{equation}
and \eqref{Metr2} can also be written as
\begin{equation}
\label{Metr3}
Q_{N }=\frac{1}{N}\sum_{k=1}^{N}T_{k}
\end{equation}

\subsection{Proposed Algorithm and System Architecture}

\begin{figure}[!t]
\centering
\includegraphics[width=3.4in]{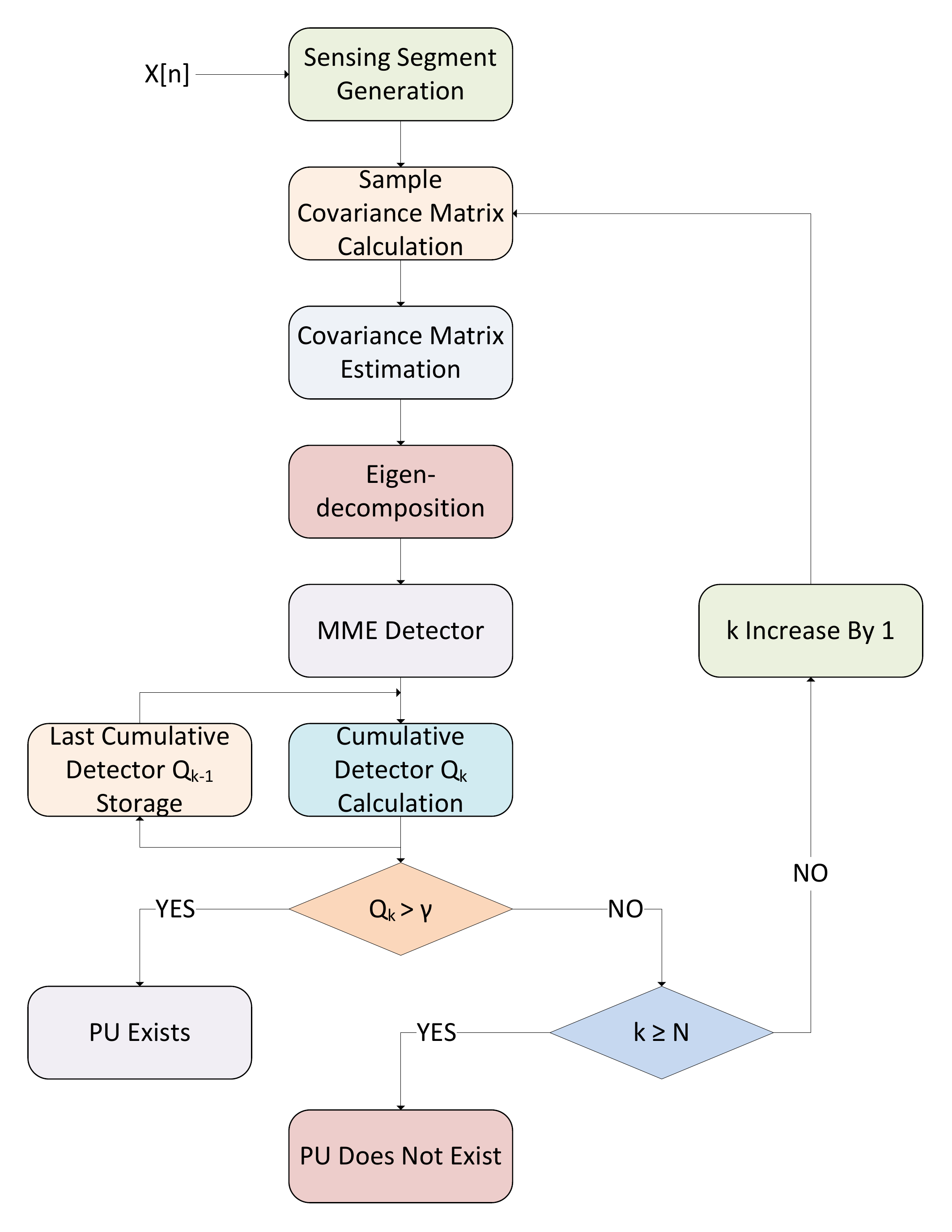}
\caption{Overall processing architecture and data flow diagram of the proposed  cumulative spectrum sensing approach}
\label{architecture}
\end{figure}

Base on the above analysis, we propose the cumulative spectrum sensing approach with small-sized datasets, which is summarized as \textbf{Algorithm 1}. Meanwhile, the overall processing module architecture and the data flow diagram are shown in Fig.~\ref{architecture}. A sequence of data are streamed into the system, every new coming sample will generate a new detection metric value, until the decision is made. The data streaming and detection are working simultaneously. The minimal required number of total data size is $L$, so that at least one multivariate vector is contained in the sensing segment.
\algsetup{indent=2em}
\begin{algorithm}
\caption{Cumulative Spectrum Sensing With Small-Sized Datasets}
\label{qsd}
\begin{algorithmic}[1]
  \STATE $\gamma \leftarrow$ Set the detection threshold 
	\STATE $\Gamma _{x,i} \leftarrow$ Sensing segment
	\STATE Initial:
	\STATE $k = 1, q_{0} = 0$
	\WHILE{True}
	\label{cal1}
	\STATE $\hat{\mathbf{R}}_{x,k} \leftarrow$ Calculate sample covariance matrix using Eq.~\eqref{cov_mat3} where $N$ is set to be $k$
	\STATE $\hat{\mathbf{F}_{k}}, \hat{\rho} _{\mathrm{OAS}_{k}} \leftarrow \hat{\mathbf{R}}_{x,k}$
	\STATE $\hat{\mathbf{\Sigma }}_{\mathrm{OAS,k}} \leftarrow \hat{\mathbf{R}}_{x,k}, \hat{\mathbf{F}_{k}}, \hat{\rho} _{\mathrm{OAS}_{k}}$
	\STATE $\lambda_{1,k}, \lambda_{L,k} \leftarrow \hat{\mathbf{\Sigma }}_{\mathrm{OAS,k}}$ // Eigen-decomposition of $\hat{\mathbf{\Sigma }}_{\mathrm{OAS,k}}$
	\STATE $T_{k} \leftarrow \lambda_{1,k}/\lambda_{L,k}$
	\STATE $q_{k} \leftarrow q_{k-1}, T_{k}$
	\STATE $Q_{k} \leftarrow q_{k}/k$
	\IF{$Q_{k} > \gamma$} 
	   \STATE PU exists and SU vacates the channel
		 \RETURN
	\ELSIF{$k \geq  N$}
	   \STATE PU does not exist
		 \RETURN
	\ELSE
	   \STATE $k = k + 1$
	\ENDIF
	\ENDWHILE

\end{algorithmic}
\end{algorithm}

The advantage of Algorithm 1 is twofold. First, in a specific situation where the environment is not so harsh, the number of stopping total data sample $N_{tot}^{stop}$ is sufficient to detect the PU promptly once the threshold is reached. The goal is to detect as quickly as possible with minimal delay, which is particularly useful for detection in vehicular applications. Second, if we use all the received data for the detection decision, the proposed algorithm is able to work in a relatively low SNR environment. Three essential properties of this algorithm are worth mentioning here:
\begin{enumerate}
    \item For detection under low SNR environment, more data brings better performance.
		\item The threshold is robust which is not related to the noise power. 
		\item The algorithm is blind without any knowledge of the signal or the noise.
\end{enumerate}

Both covariance matrix estimation and cumulative iteration contribute to the proposed algorithm. In order to show how much each part contributes in this Algorithm 1, we propose an additional \textbf{Algorithm 2} which merely involves covariance matrix estimation. The initial $k$ is set to be $N$; $T_{N}$ then can be obtained as before, after that $T_{N}$ is directly compared with the threshold $\gamma$ to make a decision. The performance of Algorithm 2 will be provided in Section~\ref{simu} as well as a reference.

\section{Performance Analysis}
\label{analy}

\subsection{Concentration Inequalities of Statistics}
The concentration inequalities of statistics will be analyzed in this section for the proposed algorithm. The detailed proof will be given based on the following (simplified version of the) theorem by McDiarmid~\cite{mcdiarmid1989method, ozgur2013spatial}.

\begin{theorem}
\label{theorem2}
Let $\mathbf{x}_{1},\cdots ,\mathbf{x}_{N}$ be independent random variables taking values in a set $A$, and let $f:A^{N}\rightarrow \mathbb{R}$ be a measurable function such that these is a constant $c$ with

$\left | f\left ( \mathbf{x}_{1},\cdots ,\mathbf{x}_{m},\cdots ,\mathbf{x}_{N} \right ) - f\left ( \mathbf{x}_{1},\cdots ,{\mathbf{x}_{m}}',\cdots ,\mathbf{x}_{N} \right )\right |\leq c$ for all $1\leq m\leq N$, $\mathbf{x}_{1},\cdots, \mathbf{x}_{m},{\mathbf{x}_{m}}',\cdots ,\mathbf{x}_{N}\in A$, and the sequence $\mathbf{x}_{1},\cdots ,\mathbf{x}_{m},\cdots ,\mathbf{x}_{N}$ and $\mathbf{x}_{1},\cdots ,{\mathbf{x}_{m}}',\cdots ,\mathbf{x}_{N}$ differ only in the $m$th co-ordinate. Then for all $t> 0$,

\begin{equation}
\label{theo}
\begin{gathered}
  {\text{P}}\left( {\left| {f\left( {{{\mathbf{x}}_1}, \cdots ,{{\mathbf{x}}_N}} \right) - {\text{E}}\left( {f\left( {{{\mathbf{x}}_1}, \cdots ,{{\mathbf{x}}_N}} \right)} \right)} \right| \geqslant t} \right) \hfill \\
   \leqslant 2{\text{exp}}\left( { - \frac{{2{t^2}}}{{N{c^2}}}} \right) \hfill \\ 
\end{gathered}
\end{equation}
\end{theorem}

Let sample covariance matrix based on sequence $\mathbf{x}_{1},\cdots ,{\mathbf{x}_{m}}',\cdots ,\mathbf{x}_{N}$ be defined as
\begin{equation}
\label{cov_mat_alte}
{\hat{\mathbf{R}}_{x,m}}'=\frac{1}{N}\left (\sum_{i=1,i\neq m}^{N}\mathbf{x}_{i}\mathbf{x}_{i}^{T} + {\mathbf{x}_{m}}'{\mathbf{x}_{m}}'^{T} \right )
\end{equation}
where $m = 1, \cdots, N$.

Accordingly, the OAS estimated covariance matrix is obtained as
\begin{equation}
\label{estcov2}
{\hat{\mathbf{\Sigma }}_{\mathrm{OAS}}}'=\left (1-{\hat{\rho}_{\mathrm{OAS}}}'   \right ) {\hat{\mathbf{R}}_{x,m}}' + {\hat{\rho}_{\mathrm{OAS}}}' {\hat{\mathbf{F}}}'
\end{equation}

The above equation can be treated as a matrix perturbation to the original OAS estimated covariance matrix, written as
\begin{equation}
\label{pertur}
{\hat{\mathbf{\Sigma }}_{\mathrm{OAS}}}' = \hat{\mathbf{\Sigma }}_{\mathrm{OAS}} + \mathbf{E}
\end{equation}
where $\mathbf{E}$, which is Hermitian, is a perturbation matrix to $\hat{\mathbf{\Sigma }}_{\mathrm{OAS}}$, .

The eigenvalues obtained from ${\hat{\mathbf{\Sigma }}_{\mathrm{OAS}}}'$ and $\mathbf{E}$ are ${\lambda _{1}}'\geq \cdots \geq {\lambda _{L}}'$ and $\lambda _{1}\left ( \boldsymbol{\mathbf{E}} \right )\geq \cdots \geq \lambda _{L}\left ( \mathbf{E} \right )$, respectively.

\begin{theorem}~\cite[p.~34]{bhatia1987perturbation} 
\label{theorem3}
Let $A, B$ be Hermitian matrices with eigenvalues $\lambda _{1}\left ( \mathbf{A} \right )\geq \cdots \geq \lambda _{L}\left ( \mathbf{A} \right )$ and $\lambda _{1}\left ( \mathbf{B} \right )\geq \cdots \geq \lambda _{L}\left ( \mathbf{B} \right )$, respectively. Then,
\begin{equation}
\label{theo2}
\lambda _{j}\left ( \mathbf{A} \right )+\lambda _{L}\left ( \mathbf{B} \right )\leq \lambda _{j}\left ( \mathbf{A+B} \right )\leq \lambda _{j}\left ( \mathbf{A} \right )+\lambda _{1}\left ( \mathbf{B} \right )
\end{equation}
\end{theorem}

\begin{theorem}~\cite[p.~34]{bhatia1987perturbation}
\label{theorem4} 
Let $A, B$ be Hermitian matrices with eigenvalues $\lambda _{1}\left ( \mathbf{A} \right )\geq \cdots \geq \lambda _{L}\left ( \mathbf{A} \right )$ and $\lambda _{1}\left ( \mathbf{B} \right )\geq \cdots \geq \lambda _{L}\left ( \mathbf{B} \right )$, respectively. Then,
\begin{equation}
\label{theo4}
\mathop {\max }\limits_j \left| {{\lambda _j}\left( {\mathbf{A}} \right) - {\lambda _j}\left( {\mathbf{B}} \right)} \right| \leqslant \left\| {{\mathbf{A}} - {\mathbf{B}}} \right\|
\end{equation}
where $\left\| {{\mathbf{A}} - {\mathbf{B}}} \right\|=\max \left\{ {\left| {{\lambda _1}\left( {{\mathbf{A}} - {\mathbf{B}}} \right)} \right|,\left| {{\lambda _L}\left( {{\mathbf{A}} - {\mathbf{B}}} \right)} \right|} \right\}$
\end{theorem}

Since $\hat{\mathbf{\Sigma }}_{\mathrm{OAS}}$ and $\mathbf{E}$ are both Hermitian, substituting $j$ with $1$ and $L$ into the inequality of Theorem~\ref{theorem3} leads to the following results:
\begin{equation}
\label{theo3}
\lambda _{1}+\lambda _{L}\left ( \mathbf{E} \right )\leq {\lambda _{1}}'
\leq \lambda _{1}+\lambda _{1}\left ( \mathbf{E} \right )
\end{equation}
\begin{equation}
\label{theo5}
\lambda _{L}+\lambda _{L}\left ( \mathbf{E} \right )\leq {\lambda _{L}}'
\leq \lambda _{L}+\lambda _{1}\left ( \mathbf{E} \right )
\end{equation}
Remember now that
\begin{equation}
\label{f_metric}
f\left ( \mathbf{x}_{1},\cdots ,\mathbf{x}_{m},\cdots ,\mathbf{x}_{N} \right )=\frac{1}{N}\sum_{k=1}^{N}\left (\frac{\lambda _{1,k}}{\lambda _{L,k}}  \right )
\end{equation}

Defining
\begin{equation}
\label{f_metric2}
f\left ( \mathbf{x}_{1},\cdots ,{\mathbf{x}_{m}}',\cdots ,\mathbf{x}_{N} \right )=\frac{1}{N}\sum_{k=1}^{N}\left (\frac{{\lambda _{1,k}}'}{{\lambda _{L,k}}'}  \right )
\end{equation}

Then,
\begin{equation}
\label{inequ}
\begin{gathered}
  \left| {f\left( {{{\mathbf{x}}_1}, \cdots ,{{\mathbf{x}}_m}, \cdots ,{{\mathbf{x}}_N}} \right) - f\left( {{{\mathbf{x}}_1}, \cdots ,{{\mathbf{x}}_m}', \cdots ,{{\mathbf{x}}_N}} \right)} \right| \hfill \\
   = \left| {\frac{1}{N}\sum\limits_{k = 1}^N {\left( {\frac{{{\lambda _{1,k}}}}{{{\lambda _{L,k}}}}} \right)}  - \frac{1}{N}\sum\limits_{k = 1}^N {\left( {\frac{{{\lambda _{1,k}}^\prime }}{{{\lambda _{L,k}}^\prime }}} \right)} } \right| \hfill \\
   = \frac{1}{N}\left| {\sum\limits_{k = 1}^N {\left( {\frac{{{\lambda _{1,k}}}}{{{\lambda _{L,k}}}} - \frac{{{\lambda _{1,k}}^\prime }}{{{\lambda _{L,k}}^\prime }}} \right)} } \right| \hfill \\
   \leqslant \frac{1}{N}\sum\limits_{k = 1}^N {\left| {\frac{{{\lambda _{1,k}}}}{{{\lambda _{L,k}}}} - \frac{{{\lambda _{1,k}}^\prime }}{{{\lambda _{L,k}}^\prime }}} \right|}  \hfill \\
   \leqslant \mathop {\max }\limits_k \left\{ {\left| {\frac{{{\lambda _{1,k}}}}{{{\lambda _{L,k}}}} - \frac{{{\lambda _{1,k}}^\prime }}{{{\lambda _{L,k}}^\prime }}} \right|} \right\} \hfill \\ 
\end{gathered} 
\end{equation}

\textit{Case 1}: ${\frac{{{\lambda _{1,k}}}}{{{\lambda _{L,k}}}} \geq \frac{{{\lambda _{1,k}}^\prime }}{{{\lambda _{L,k}}^\prime }}}$

\begin{equation}
\label{inequ2}
\begin{gathered}
  \left| {\frac{{{\lambda _{1,k}}}}{{{\lambda _{L,k}}}} - \frac{{{\lambda _{1,k}}^\prime }}{{{\lambda _{L,k}}^\prime }}} \right| = \frac{{{\lambda _{1,k}}}}{{{\lambda _{L,k}}}} - \frac{{{\lambda _{1,k}}^\prime }}{{{\lambda _{L,k}}^\prime }} \hfill \\
   \leqslant \frac{{{\lambda _{1,k}}}}{{{\lambda _{L,k}}}} - \frac{{{\lambda _{1,k}}}}{{{\lambda _{L,k}}^\prime }} \hfill \\
   = \frac{{{\lambda _{1,k}}\left( {{\lambda _{L,k}}^\prime  - {\lambda _{L,k}}} \right)}}{{{\lambda _{L,k}}^\prime {\lambda _{L,k}}}} \hfill \\ 
\end{gathered} 
\end{equation}
By applying Theorem~\ref{theorem4}, 
\begin{equation}
\label{inequ3}
{\lambda _{L,k}}^\prime  - {\lambda _{L,k}} \leqslant \mathop {\max }\limits_j \left| {{\lambda _{j,k}}^\prime  - {\lambda _{j,k}}} \right| \leqslant \left\| {\mathbf{E}} \right\|
\end{equation}
where $\left\| {\mathbf{E}} \right\| = \max \left\{ {\left| {{\lambda _1}\left( {\mathbf{E}} \right)} \right|,\left| {{\lambda _L}\left( {\mathbf{E}} \right)} \right|} \right\}$.
Hence
\begin{equation}
\label{inequ4}
\frac{{{\lambda _{1,k}}\left( {{\lambda _{L,k}}^\prime  - {\lambda _{L,k}}} \right)}}{{{\lambda _{L,k}}^\prime {\lambda _{L,k}}}} \leqslant \frac{{{\lambda _{1,k}}\left\| {\mathbf{E}} \right\|}}{{{\lambda _{L,k}}^\prime {\lambda _{L,k}}}}
\end{equation}

\textit{Case 2}: ${\frac{{{\lambda _{1,k}}}}{{{\lambda _{L,k}}}} < \frac{{{\lambda _{1,k}}^\prime }}{{{\lambda _{L,k}}^\prime }}}$

\begin{equation}
\label{inequ5}
\begin{gathered}
  \left| {\frac{{{\lambda _{1,k}}}}{{{\lambda _{L,k}}}} - \frac{{{\lambda _{1,k}}^\prime }}{{{\lambda _{L,k}}^\prime }}} \right| = \frac{{{\lambda _{1,k}}^\prime }}{{{\lambda _{L,k}}^\prime }} - \frac{{{\lambda _{1,k}}}}{{{\lambda _{L,k}}}} \hfill \\
   \leqslant \frac{{{\lambda _{1,k}}^\prime }}{{{\lambda _{L,k}}}} - \frac{{{\lambda _{1,k}}}}{{{\lambda _{L,k}}}} \hfill \\
   = \frac{{{\lambda _{1,k}}^\prime  - {\lambda _{1,k}}}}{{{\lambda _{L,k}}}} \hfill \\ \leqslant \frac{{\left\| {\mathbf{E}} \right\|}}{{{\lambda _{L,k}}}} \hfill \\
\end{gathered} 
\end{equation}

The result is also based on Theorem~\ref{theorem4}, similar to Case 1.
%

Consequently,
\begin{equation}
\label{inequ6}
\begin{gathered}
  \left| {f\left( {{{\mathbf{x}}_1}, \cdots ,{{\mathbf{x}}_m}, \cdots ,{{\mathbf{x}}_N}} \right) - f\left( {{{\mathbf{x}}_1}, \cdots ,{{\mathbf{x}}_m}^\prime , \cdots ,{{\mathbf{x}}_N}} \right)} \right| \hfill \\
   \leqslant \mathop {\max }\limits_k \left\{ {\frac{{\left\| {\mathbf{E}} \right\|}}{{{\lambda _{L,k}}}}\max \left( {\frac{{{\lambda _{1,k}}}}{{{\lambda _{L,k}}^\prime }},1} \right)} \right\} \hfill \\ 
\end{gathered} 
\end{equation}

Based on Theorem~\ref{theorem2}, we finally obtain that for all $t > 0$
\begin{equation}
\label{final}
\begin{gathered}
  {\text{P}}\left( {\left| {f\left( {{{\mathbf{x}}_1}, \cdots ,{{\mathbf{x}}_N}} \right) - {\text{E}}\left( {f\left( {{{\mathbf{x}}_1}, \cdots ,{{\mathbf{x}}_N}} \right)} \right)} \right| \geqslant t} \right) \hfill \\
   \leqslant 2{\text{exp}}\left( { - \frac{{2{t^2}}}{{N{c^2}}}} \right) \hfill \\ 
\end{gathered} 
\end{equation}
where $c = \mathop {\max }\limits_k \left\{ {\frac{{\left\| {\mathbf{E}} \right\|}}{{{\lambda _{L,k}}}}\max \left( {\frac{{{\lambda _{1,k}}}}{{{\lambda _{L,k}}^\prime }},1} \right)} \right\}$.

This concentration analysis shows that for a small value of $t$, with high probability
\begin{equation}
\label{concen}
\mathrm{E}\left( Q_{N} \right )-t \leq Q_{N}\leq \mathrm{E}\left ( Q_{N} \right )+t
\end{equation}
which indicates $Q_{N}$ highly concentrates at its mean value $\mathrm{E}\left ( Q_{N} \right )$. Therefore, under null hypothesis and alternative hypothesis
\begin{equation}
\label{concen2}
\mathrm{E}\left( Q_{N,0} \right )-t \leq Q_{N}\leq \mathrm{E}\left ( Q_{N,0} \right )+t
\end{equation}
\begin{equation}
\label{concen3}
\mathrm{E}\left( Q_{N,1} \right )-t \leq Q_{N}\leq \mathrm{E}\left ( Q_{N,1} \right )+t
\end{equation}
where $\mathrm{E}\left( Q_{N,0} \right )$ and $\mathrm{E}\left( Q_{N,1} \right )$ are the mean values under $\mathcal{H}_{0}$ and $\mathcal{H}_{1}$. The detection statistic is able to discriminate between these two hypotheses when $\mathrm{E}\left( Q_{N,1} \right ) > \mathrm{E}\left( Q_{N,0} \right )$, denoted as
\begin{equation}
\label{concen4}
\mathcal{H}_{0} : Q_{N}\leq \mathrm{E}\left ( Q_{N,0} \right )+t
\end{equation}
\begin{equation}
\label{concen5}
\mathcal{H}_{1} : Q_{N}\geq \mathrm{E}\left ( Q_{N,1} \right )-t
\end{equation}

\begin{figure}[!t]
\centering
\includegraphics[width=3.4in]{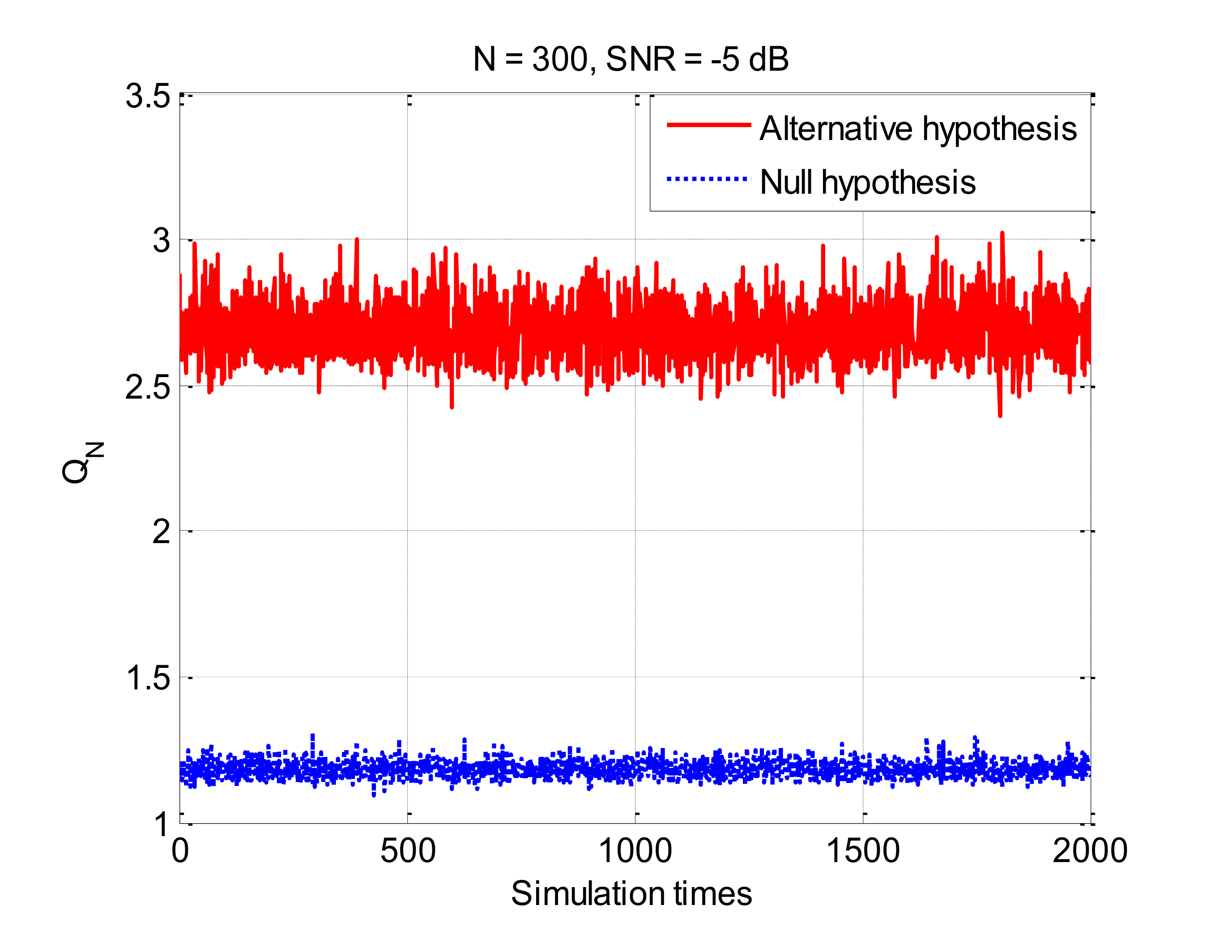}
\caption{Statistics realization distance between two hypotheses}
\label{distance1}
\end{figure}

Two thousand Monte-Carlo simulations for statistics under both hypotheses are shown in Fig.~\ref{distance1}, with SNR = -5 dB and $N = 300$. Fig.~\ref{distance1} shows that the statistics concentrate at two different values under null and alternative hypotheses, meanwhile $\mathrm{E}\left( Q_{N,1} \right )$ is greater than $\mathrm{E}\left( Q_{N,0} \right )$.

\subsection{Robust Threshold}
Generally, we have no information on the signal, it is difficult to set the threshold based on the $P_{d}$. Thus, we choose the threshold based on $P_{fa}$. In our proposed algorithm, the sample covariance matrix is obtained with very limited samples which is inaccurate to true covariance matrix. After the covariance matrix estimation, the effect of OAS estimated covariance matrix $\hat{\mathbf{\Sigma }}_{\mathrm{OAS}}$ is equivalent to sample covariance matrix $\hat{\mathbf{R}}_{x}\left ( N_{eq} \right )$ with large number of samples, denoted as $N_{eq}$, though we don't know exactly how large $N_{eq}$ is. Hence, we can use the distribution of eigenvalues obtained from $\hat{\mathbf{R}}_{x}\left ( N_{eq} \right )$ to approximate the distribution of eigenvalues obtained from $\hat{\mathbf{\Sigma }}_{\mathrm{OAS}}$. 

\begin{theorem}~\cite{johnstone2001distribution, 5089517} 
\label{theorem5}
Assume that noise is real. Let $\mathbf{A}\left ( N_{eq} \right )=\frac{N_{eq}}{\sigma ^{2}}\hat{\mathbf{R}}_{x}\left ( N_{eq} \right )$, $\mu = \left ( \sqrt{N_{eq}-1} + \sqrt{L}\right )^{2}$ and $\nu  = \left ( \sqrt{N_{eq}-1} + \sqrt{L}\right )\left ( \frac{1}{\sqrt{N_{eq}-1}} +\frac{1}{\sqrt{L}}  \right )^{\frac{1}{3}}$. Then, $\frac{\lambda _{1}\left ( \mathbf{A}\left ( N_{eq} \right ) \right )-\mu }{\nu }$ converges to the Tracy-Widom distribution of order 1 $(W_{1})$.
\end{theorem}

The mean and variance of Tracy-Widom distribution of order 1 can be found~\cite{bornemann2010numerical} to be $\mu _{tw}=-1.20653$ and $\sigma_{tw}^{2}=1.60778$. It's easy to obtain the mean and variance of $\lambda _{1}\left ( \mathbf{A}\left ( N_{eq} \right ) \right )$ as $\mu + \nu \mu _{tw}$ and $\nu ^{2}\sigma _{tw}^{2}$, respectively. And hence, the the mean and variance of $\lambda _{1}\left ( \hat{\mathbf{R}}_{x}\left ( N_{eq} \right ) \right )$ to be $\frac{\sigma ^{2}\left ( \mu + \nu \mu _{tw} \right )}{N_{eq}}$ and $\frac{\sigma ^{4}\nu ^{2}\sigma _{tw}^{2}}{N_{eq}^{2}}$, respectively.

\begin{theorem}~\cite{bai1999methodologies, 5089517}
\label{theorem6}
Assume that $\mathop {\lim }\limits_{{N_{eq}} \to \infty } \frac{L}{{{N_{eq}}}} = y\left( {0 < y < 1} \right)$. Then, $\mathop {\lim }\limits_{{N_{eq}} \to \infty } {\lambda _{L}\left ( \hat{\mathbf{R}}_{x}\left ( N_{eq} \right ) \right )} = {\sigma ^2}{\left( {1 - \sqrt y } \right)^2}$ (with probability one).
\end{theorem}

Based on the Theorem~\ref{theorem6}, the smallest eigenvalue of $\hat{\mathbf{R}}_{x}\left ( N_{eq} \right )$ tend to be deterministic value $\frac{{\sigma ^2}}{N_{eq}}{\left( {\sqrt{N_{eq}} - \sqrt L } \right)^2}$ when $N_{eq}$ is large. In such case, $\frac{\lambda _{1}\left ( \hat{\mathbf{R}}_{x}\left ( N_{eq} \right ) \right )}{{\lambda _{L}\left ( \hat{\mathbf{R}}_{x}\left ( N_{eq} \right ) \right )}}$ can be viewed as a new random variable $T$ obtained from random variable $\lambda _{1}\left ( \hat{\mathbf{R}}_{x}\left ( N_{eq} \right ) \right )$ with a coefficient $\frac{1}{\lambda _{L}\left ( \hat{\mathbf{R}}_{x}\left ( N_{eq} \right ) \right )}$.

As a result, the mean and variance of $T$ are written as
\begin{equation}
\label{mean_metri}
\begin{gathered}
  \mu_{T} =  \frac{{{{\left( {\sqrt {{N_{eq}} - 1}  + \sqrt L } \right)}^2}}}{{{{\left( {\sqrt {{N_{eq}}}  - \sqrt L } \right)}^2}}} \hfill \\
   + \tfrac{{\left( {\sqrt {{N_{eq}} - 1}  + \sqrt L } \right){{\left( {\frac{1}{{\sqrt {{N_{eq}} - 1} }} + \frac{1}{{\sqrt L }}} \right)}^{\frac{1}{3}}}{\mu _{tw}}}}{{{{\left( {\sqrt {{N_{eq}}}  - \sqrt L } \right)}^2}}} \hfill \\ 
\end{gathered} 
\end{equation}

\begin{equation}
\label{var_metri}
\begin{gathered}
  \sigma _{T}^{2}  = \frac{{{{\left( {\sqrt {{N_{eq}} - 1}  + \sqrt L } \right)}^2}{{\left( {\frac{1}{{\sqrt {{N_{eq}} - 1} }} + \frac{1}{{\sqrt L }}} \right)}^{\frac{2}{3}}}\sigma _{tw}^2}}{{{{\left( {\sqrt {{N_{eq}}}  - \sqrt L } \right)}^4}}} \hfill \\ 
\end{gathered} 
\end{equation}

The final detection statistic $Q_{N}$ is the arithmetic average of $T$, based on central limit theorem, $Q_{N}$ follows Gaussian distribution $\mathcal{N}\left ( \mu_{Q_{N}}, \sigma _{Q_{N}}^{2} \right )$ with mean and variance as follows:

\begin{equation}
\label{mean_metri2}
\begin{gathered}
  \mu_{Q_{N}} =   \frac{{{{\left( {\sqrt {{N_{eq}} - 1}  + \sqrt L } \right)}^2}}}{{{{\left( {\sqrt {{N_{eq}}}  - \sqrt L } \right)}^2}}} \hfill \\
   + \tfrac{{\left( {\sqrt {{N_{eq}} - 1}  + \sqrt L } \right){{\left( {\frac{1}{{\sqrt {{N_{eq}} - 1} }} + \frac{1}{{\sqrt L }}} \right)}^{\frac{1}{3}}}{\mu _{tw}}}}{{{{\left( {\sqrt {{N_{eq}}}  - \sqrt L } \right)}^2}}} \hfill \\ 
\end{gathered} 
\end{equation}

\begin{equation}
\label{var_metri2}
\begin{gathered}
  \sigma _{Q_{N}}^{2}   = \frac{{{{\left( {\sqrt {{N_{eq}} - 1}  + \sqrt L } \right)}^2}{{\left( {\frac{1}{{\sqrt {{N_{eq}} - 1} }} + \frac{1}{{\sqrt L }}} \right)}^{\frac{2}{3}}}\sigma _{tw}^2}}{{{{N\left( {\sqrt {{N_{eq}}}  - \sqrt L } \right)}^4}}} \hfill \\ 
\end{gathered} 
\end{equation}

\begin{figure}[!t]
\centering
\includegraphics[width=3.2in]{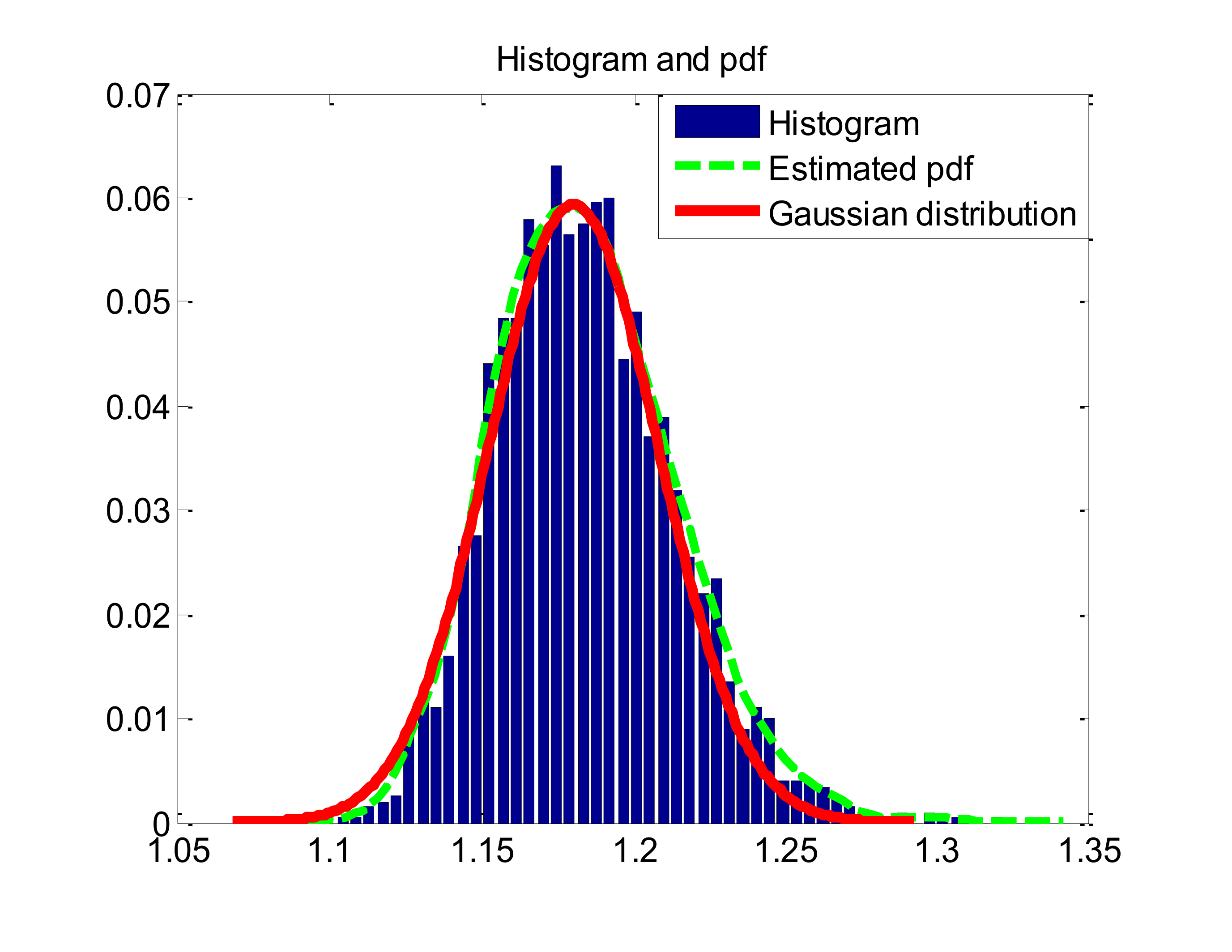}
\caption{Histogram and pdf of statistic under null hypothesis when N = 300, SNR = -10 dB}
\label{noisepdf2}
\end{figure}

By taking advantage of covariance matrix estimation, multiple (i.e., $N$) random variables $T$ are generated provided limited total samples (i.e., $N + L -1$). Because of the cumulative average, the variance of the random variable can be further reduced by a factor of $N$ to reach \eqref{var_metri2}. The histogram and the estimated probability distribution function (pdf) of the statistic under $\mathcal{H}_{0}$ are shown in Fig.~\ref{noisepdf2}. We can see the pdf approximates a Gaussian distribution.

The false alarm probability can be transformed into standard Gaussian distribution form as
\begin{equation}
\label{gaussian}
\begin{gathered}
  {P_{fa}} = P\left( {\frac{{{Q_N} - {\mu _{{Q_N}}}}}{{{\sigma _{{Q_N}}}}} > \frac{{\gamma  - {\mu _{{Q_N}}}}}{{{\sigma _{{Q_N}}}}}} \right) \hfill \\
  \quad \quad  = {\text{Q}}\left( {\frac{{\gamma  - {\mu _{{Q_N}}}}}{{{\sigma _{{Q_N}}}}}} \right) \hfill \\ 
\end{gathered} 
\end{equation}
where
\begin{equation}
\label{Qf}
\textrm{Q}\left ( t \right )=\frac{1}{\sqrt{2\pi }}\int_{t}^{+\infty }e^{\frac{-x^{2}}{2}}\textrm{d}x
\end{equation}
then the threshold $\gamma$ can be calculated as
\begin{equation}
\label{Qf2}
\gamma =\mu _{Q_{N}}+\textrm{Q}^{-1}\left ( P_{fa} \right )\sigma _{Q_{N}}
\end{equation}

Here the threshold is not affected by noise power or SNR, which is stable against environment changes.

\section{Numerical Results}
\label{simu}

In this section, we will give some simulation results using a digital TV (DTV) signal, which was captured (field measurements) in Washington D.C.~\cite{DTV2006Measurements}. The data rate of the vestigial sideband (VSB) DTV signal is 10.762 MSamples/sec. The recorded DTV signal was sampled at receiver at 21.524476 MSamples/sec and down converted to a low central IF frequency of 5.381119 MHz. The SNR of the received signal is unknown. In order to use the signals for simulating the algorithms at low SNR, we need to add white Gaussian noise to obtain various SNR levels~\cite{shellhammer2006spectrum}. The smoothing factor $L$ is chosen to be 32. False alarm probability is fixed with $P_{fa}=1\%$, and all the thresholds are determined by this 1\% false alarm probability. Two thousand simulations are performed on different sample sizes or different SNR levels. 

\subsection{Simulations on Proposed Algorithms}
\begin{figure}[!t]
\centering
\includegraphics[width=3.2in]{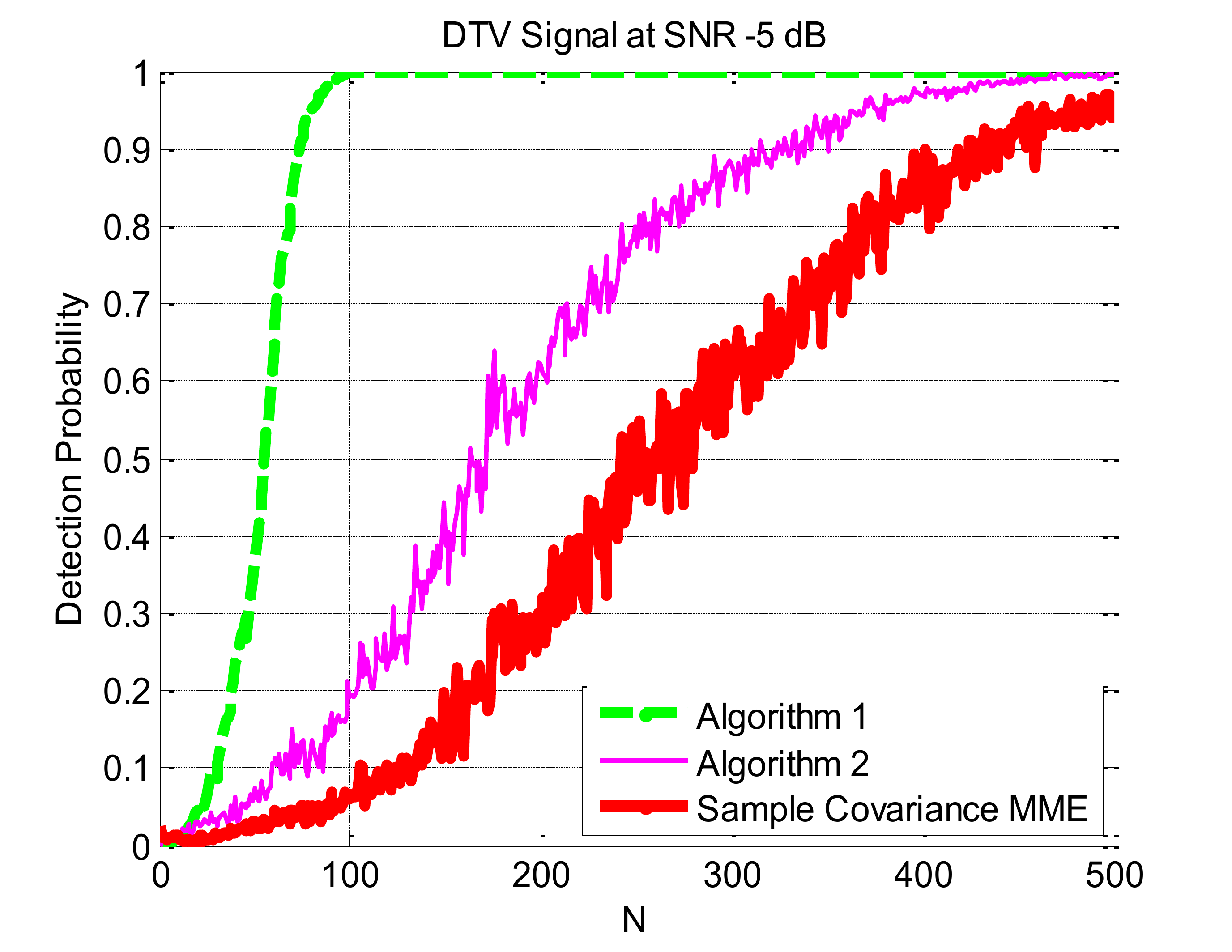}
\caption{Detection probabilities of proposed algorithms at SNR = -5 dB with DTV signal}
\label{dtv_500_minus5}
\end{figure}

\begin{figure}[!t]
\centering
\includegraphics[width=3.2in]{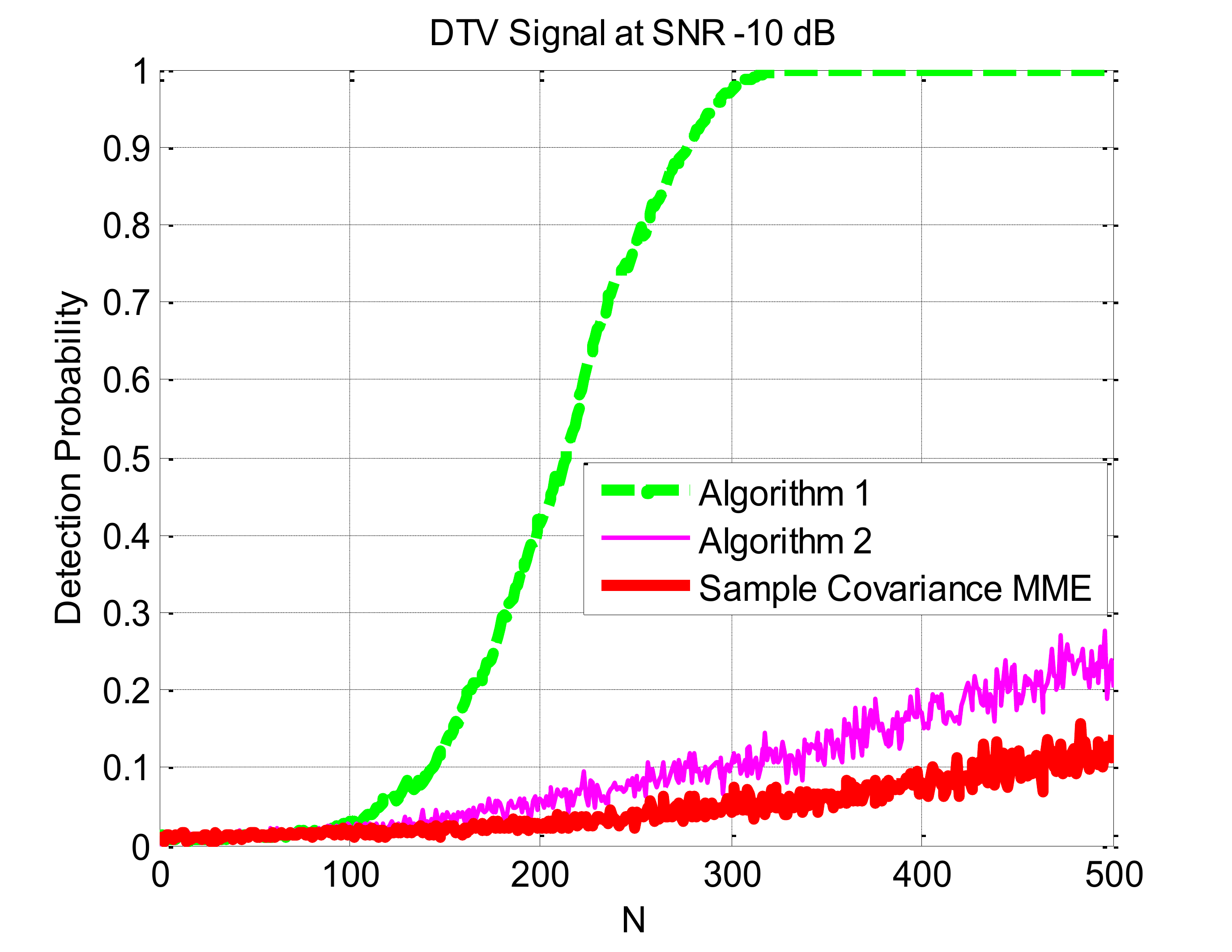}
\caption{Detection probabilities of proposed algorithms at SNR = -10 dB with DTV signal}
\label{dtv_500_minus10}
\end{figure}

In Fig.~\ref{dtv_500_minus5}, SNR is fixed at -5 dB while N varies. For 100\% detection probability, Algorithm 1 only needs a sample size of 100 to achieve that. It is equivalent to 131 total data, corresponding to 6.086 micro seconds. While Algorithm 2 needs about a sample size of 500, that is 531 total data, corresponding to 24.670 micro seconds. The original sample covariance MME requires more than a sample size of 500 to achieve the same detection probability. The performance of detection with SNR -10 dB is shown in Fig.~\ref{dtv_500_minus10}. With lower SNR, Algorithm 1 needs more data to achieve required detection probability. Observing from the figure, it's about a sample size of 320 to reach 100\% detection probability, which needs 16.307 micro seconds. The detection probabilities of both Algorithm 2 and sample covariance MME are increasing slowly as the N grows. They need far more data to obtain a satisfied detection performance when SNR is low, yet Algorithm 2 still performs better than sample covariance MME.

\begin{figure}[!t]
\centering
\includegraphics[width=3.2in]{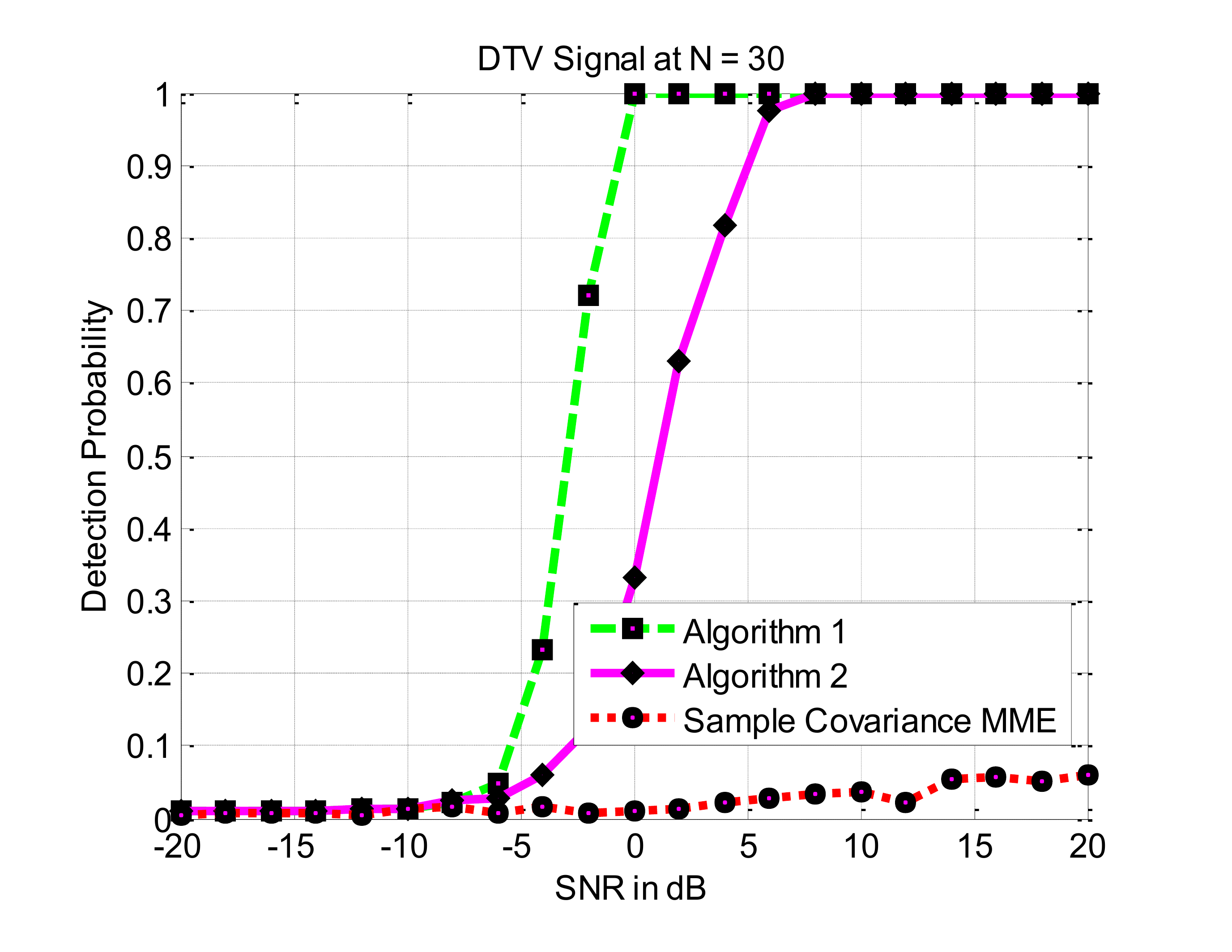}
\caption{Detection probabilities of proposed algorithms at N = 30 with DTV signal}
\label{dtv_n30}
\end{figure}

\begin{figure}[!t]
\centering
\includegraphics[width=3.2in]{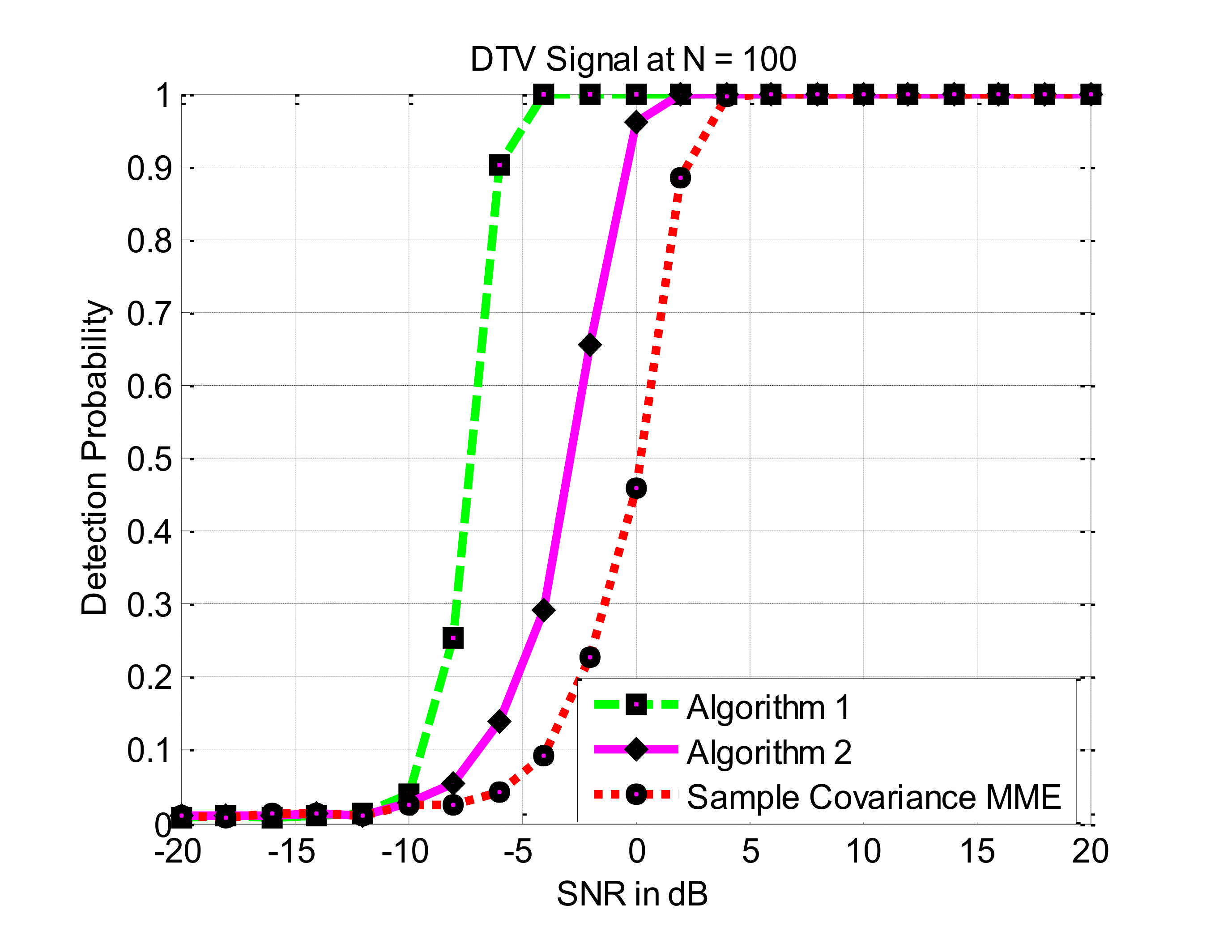}
\caption{Detection probabilities of proposed algorithms at N = 100 with DTV signal}
\label{dtv_n100}
\end{figure}

Given the number of total data size, the proposed algorithms are able to work in a relatively low SNR environment. In Fig.~\ref{dtv_n30}, when N equals 30, Algorithm 1 and Algorithm 2 can work at SNR 0 dB and 8 dB to achieve 100\% detection probability, respectively. However, sample covariance MME is ineffective at any SNR level. When N increases to 100, shown in Fig.~\ref{dtv_n100}, all the detection performance are improved. Algorithm 1 can work at -4 dB, which is a 8 dB gain compared with sample covariance MME working at 4 dB. Fig.~\ref{dtv_n30} and Fig.~\ref{dtv_n100} suggested that providing more data will help detect the signal in a lower SNR environment. The SNR and the sample size N needed to achieve 100\% detection probability exhibited a linear relationship between them, as shown in Table~\ref{table_L}. When the SNR (in dB) decreases by 3 dB, which equals SNR (not in dB) reduces by a factor of 2, the N will accordingly increase by a factor around 2.
 
\begin{table}[!t]
\renewcommand{\arraystretch}{1.2}
\caption{Linear relationship between N and SNR to achieve 100\% detection probability}
\label{table_L}
\centering
\begin{tabular}{|c||c|c|c|c|c|c|}
\hline
SNR (in dB) & 3 dB & 0 dB & -3 dB & -6 dB & -9 dB & -12 dB\\
\hline
SNR (not in dB) & 2 & 1 & 0.5 & 0.25 & 0.125 & 0.0625\\
\hline
N  & 15 & 32 & 61 & 123 & 252 & 590 \\
\hline
N increases by & / & 2.13 & 1.91 & 2.02 & 2.05 & 2.34 \\
\hline
\end{tabular}
\end{table}

\begin{figure}[!t]
\centering
\includegraphics[width=3.2in]{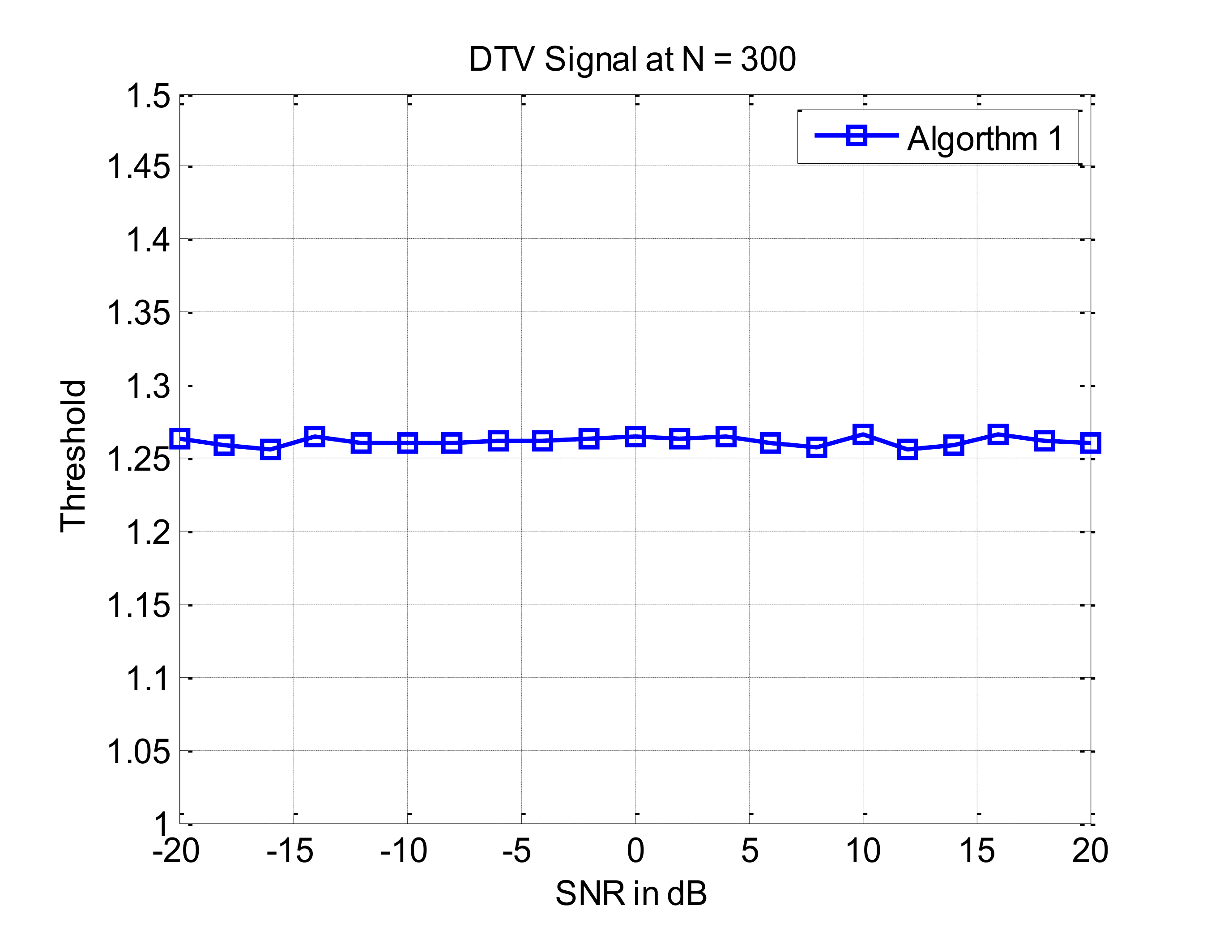}
\caption{Threshold of Algorithm 1 at N = 300 with SNR varies}
\label{Thre_n300}
\end{figure}

One of the properties of our proposed algorithm is that the threshold is robust. As shown in Fig.~\ref{Thre_n300}, the threshold is almost a constant between 1.25 and 1.3 no matter what SNR is.

\subsection{Comparison with Other Algorithms}
\label{cov}
In the following, we will discuss some other spectrum sensing algorithms for comparison purposes. Arithmetic-to-geometric mean (AGM)~\cite{4698617} which derives from GLRT, is able to sense the spectrum without prior knowledge. Feature template matching (FTM)~\cite{Zhang} utilizes feature, which can be learned blindly, as prior knowledge for detection. Estimator-correlator (EC)~\cite{dete_theo} requires both the original PU's signal covariance matrix and the noise variance. Energy detection (ED) is easy to be implemented, but usually suffers the noise uncertainty problem.

\subsubsection{AGM}
AGM finds an unstructured estimate of $\mathbf{R}_{x}$ to be $\mathbf{R}_{s}+\sigma_{n}^{2}\mathbf{I}$ under $\mathcal{H}_{1}$ and $\sigma_{n}^{2}\mathbf{I}$ under $\mathcal{H}_{0}$. $\lambda_{m}$ represents all eigenvalues of the sample covariance matrix and $M$ is the dimension of the sample covariance matrix. Since the arithmetic mean is larger than geometric mean, the resulting detector computes the arithmetic-to-geometric mean of the eigenvalues of sample covariance matrix and compares with a threshold~\cite{4698617}
\begin{equation}
\label{AGM}
T_{AGM}= \frac{\frac{1}{M}\sum_{m}\lambda_{m}}{\left (\prod_{m}\lambda_{m}  \right )^{\frac{1}{M}} }> \gamma _{AGM}
\end{equation}

\subsubsection{FTM}
FTM extracts leading eigenvector as the feature, which is stable for signals and random for noise. FTM involves two steps. First, it learns the feature blindly by comparing the similarity of two consecutive sensing segments, namely feature learning algorithm (FLA)~\cite{Zhang}.
\begin{equation}
\label{FLA}
T_{FLA}  = \mathop {\max }\limits_{l = 1,2, \cdots L - k + 1} \left| {\sum\limits_{k = 1}^L {\eta _i \left[ k \right]\eta _{i + 1} \left[ {k + l} \right]} } \right|
\end{equation}
If $T_{FLA} > \gamma_{e}$ feature is learned as $\phi _{s,1}=\eta _{i+1}$. $\gamma_{e}$ is the threshold determined empirically. Then, with the learned signal feature $\phi _{s,1}$ as prior knowledge, this algorithm compares just the feature $\phi _{x,1}$ from new sensing segment and signal feature $\phi _{s,1}$ to determine if the signal exists.
\begin{equation}
\label{FTM}
T_{FTM}  = \mathop {\max }\limits_{l = 1,2, \cdots L - k + 1} \left| {\sum\limits_{k = 1}^L {\phi _{s,1} \left[ k \right]\phi _{x,1} \left[ {k + l} \right]} } \right| > \gamma _{FTM} 
\end{equation}

\subsubsection{EC}
EC method assumes the signal follows zero mean Gaussian distribution with covariance matrix $\mathbf{R}_{s}$, and noise follows zero mean Gaussian distribution with covariance matrix $\sigma _{n}^{2}\mathbf{I}$,
\begin{equation}
\label{EC1}
s\sim \mathcal{N}\left ( 0,\mathbf{R}_{s} \right )
\end{equation}
\begin{equation}
\label{EC2}
w\sim \mathcal{N}\left ( 0,\sigma _{n}^{2}\mathbf{I} \right )  
\end{equation}
Both $\mathbf{R}_{s}$ and $\sigma_{n}^{2}$ are given priorly. The hypothesis $\mathcal{H}_{1}$ is true if~\cite{dete_theo}
\begin{equation}
\label{EC3}
T_{EC}= \sum_{j=1}^{N}\mathbf{x}_{j}^{T}\mathbf{R}_{s}\left (\mathbf{R}_{s}+\sigma _{n}^{2}\mathbf{I } \right )^{-1}\mathbf{x}_{j}>\gamma _{EC}
\end{equation}


\subsubsection{Comparison}
\begin{figure}[!t]
\centering
\includegraphics[width=3.2in]{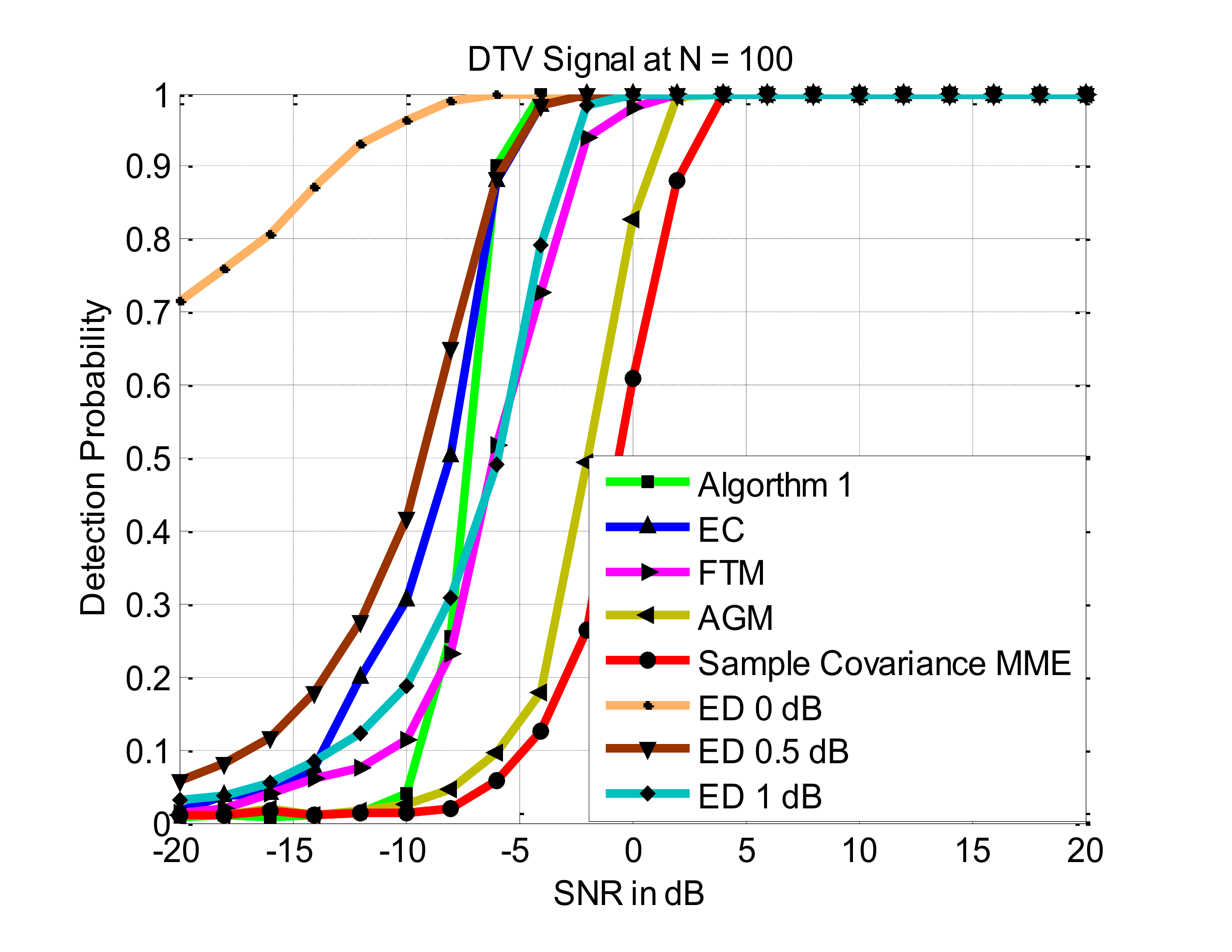}
\caption{Detection probabilities comparison among different algorithms at N = 100 with SNR varies}
\label{dtv_100_all6}
\end{figure}

\begin{figure}[!t]
\centering
\includegraphics[width=3.2in]{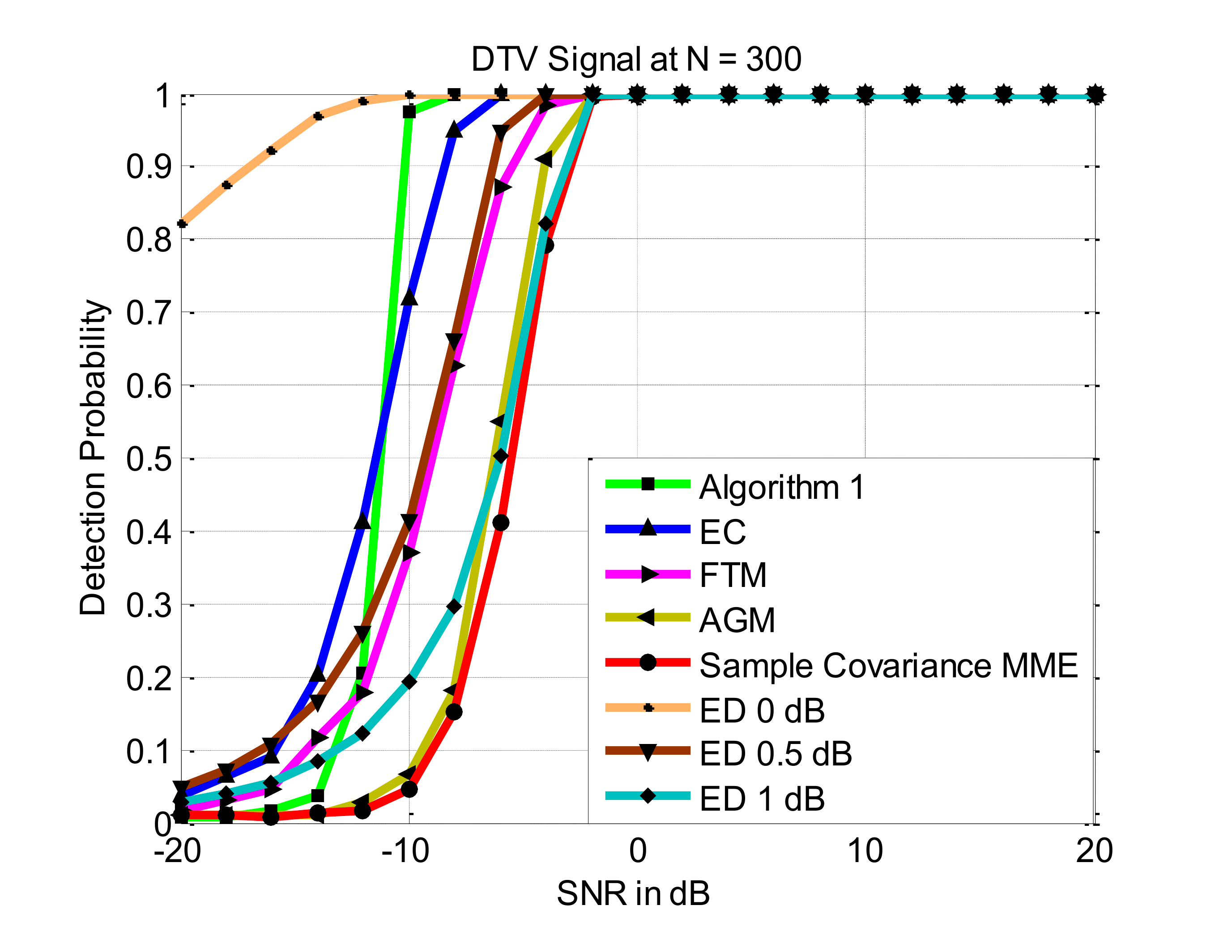}
\caption{Detection probabilities comparison among different algorithms at N = 300 with SNR varies}
\label{dtv_300_all6}
\end{figure}

\begin{figure}[!t]
\centering
\includegraphics[width=3.2in]{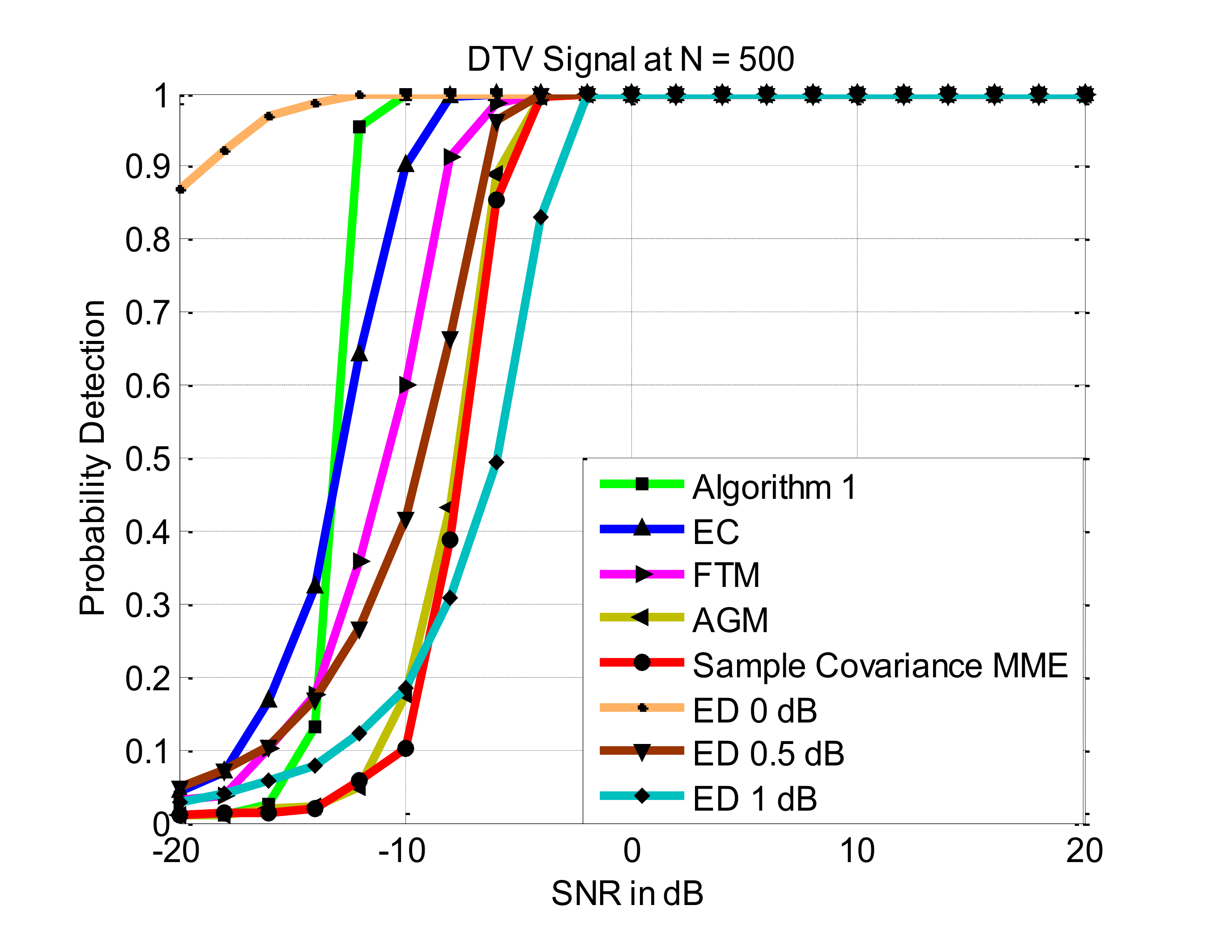}
\caption{Detection probabilities comparison among different algorithms at N = 500 with SNR varies}
\label{dtv_500_all6}
\end{figure}

The detection probabilities varied by SNR for Algorithm 1 compared with EC, FTM, AGM, MME and ED are shown in the following, where ``ED x dB'' means the energy detection with x dB noise uncertainty. In Fig.~\ref{dtv_100_all6}, when N = 100, if the noise variance is exactly known (ED 0 dB), the energy detection is pretty good. Speaking of more than 80\% detection probability, Algorithm 1 exhibits almost the same performance with EC and ED 0.5 dB, even a little better. The rest of the algorithms including FTM, AGM, MME, ED 1 dB, all require higher SNR to achieve the same detection probability. In Fig.~\ref{dtv_300_all6} when N = 300 and Fig.~\ref{dtv_500_all6} when N = 500, Algorithm 1 is only worse than ED 0 dB when considering a more than 60\% detection probability. In a word, the proposed algorithm demonstrates superior advantage, almost approximates the optimal EC and outperforms the performance of the rest of the algorithms mentioned above.

\subsubsection{Discussion}
The SNR change brought some impact to calculate $T_{\tau }$ with each sample size $\tau$, $\tau = 1,\cdots , N$. Because of the cumulative addition, this impact from SNR was accumulated and amplified with N times in forming $Q_{N}$ in alternative hypothesis, while $Q_{N}$ in null hypothesis remained unchanged. That is the reason why our algorithm is sensitive to the SNR, as shown in Fig.~\ref{dtv_100_all6} that the curve of Algorithm 1 increased steeply when SNR changes between -10 dB and -4 dB. The slope will be even steeper when the sample size N increases, as shown in Fig.~\ref{dtv_300_all6} and Fig.~\ref{dtv_500_all6}.

\section{Conclusion}
\label{conc}
We have considered the spectrum sensing for single PU with single antenna. A blind cumulative spectrum sensing method has been proposed focusing on small-sized datasets. If the total sample size is given, this method works in a lower SNR environment compared with some other algorithms. Concentration inequalities of statistics have been adopted to demonstrate the effectiveness and validity of the proposed method. Meanwhile, the threshold has been proven to be stable. All the results were verified by the simulations using a captured DTV signal.

This proposed method can also be extended to be a general detection framework. The MME detector in this method can be replaced by other covariance matrix based spectrum sensing algorithms~\cite{milcomfeng2012}, depends on different detection scenarios, the detection performance may be further improved consequently. In addition, this method can be applied to Smart Grid as well because real-time response of system changes is also a fundamental requirement in Smart Grid~\cite{smartgridbook2004}.

%


\section*{Acknowledgment}
The authors would like to thank Dr. Zhen Hu, for the helpful discussions on this paper.
\ifCLASSOPTIONcaptionsoff
  \newpage
\fi


\bibliographystyle{IEEEtran}
\bibliography{bib/library,bib/Eric}

\end{document}